\documentclass[useAMS, usenatbib, onecolumn]{mn2e}
\usepackage{graphicx}
\usepackage{amssymb}

\newcommand\be{\begin{equation}}
\newcommand\ee{\end{equation}}
\newcommand\ba{\begin{eqnarray}}
\newcommand\ea{\end{eqnarray}}

\newcommand\bb[1] {\mbox{\boldmath{$#1$}}}
\newcommand{\Alfven}{Alfv\'{e}n~}
\newcommand\tomega{{\tilde\omega}}

\def\go{\mathrel{\raise.3ex\hbox{$>$}\mkern-14mu
             \lower0.6ex\hbox{$\sim$}}}
\def\lo{\mathrel{\raise.3ex\hbox{$<$}\mkern-14mu
             \lower0.6ex\hbox{$\sim$}}}

\title[Corotational instability in magnetized discs]
{Corotational Instability, Magnetic Resonances and Global
Inertial-Acoustic Oscillations in Magnetized Black-Hole Accretion Discs}

\author[W.~Fu \& D.~Lai]
{Wen Fu$^1$\thanks{Email: wenfu@astro.cornell.edu;
dong@astro.cornell.edu} and Dong Lai$^{1,2}$\footnotemark[1]\\
$^1$Department of Astronomy, Cornell University, Ithaca, NY 14853, USA\\
$^2$KITP, University of California, Santa Barbara, CA 93106, USA\\}

\pagerange{\pageref{firstpage}--\pageref{lastpage}} \pubyear{2010}

\begin{document}
\label{firstpage}
\maketitle

\begin{abstract}

Low-order, non-axisymmetric p-modes (also referred as
inertial-acoustic modes) in hydrodynamic accretion discs
around black holes are plausible candidates for high-frequency
quasi-periodic oscillations (QPOs) observed in a number of accreting
black-hole systems. These modes are trapped in the inner-most region
of the accretion disc, and are subject to global instabilities due to
wave absorption at the corotation resonance (where the wave pattern
frequency $\omega/m$ equals the disc rotation rate $\Omega$), when the
fluid vortensity, $\zeta=\kappa^2/(2\Omega\Sigma)$ (where $\kappa$ and
$\Sigma$ are the radial epicyclic frequency and disc surface density,
respectively), has a positive gradient. We investigate the effects of
disc magnetic fields on the wave absorption at corotation and the
related wave super-reflection of the corotation barrier, and on the
overstability of disc p-modes. In general, in the presence of magnetic
fields, the p-modes have the character of inertial-fast magnetosonic
waves in their propagation zone. For discs with a pure toroidal field,
the corotation resonance is split into two magnetic resonances, where
the wave frequency in the corotating frame of the fluid,
$\tomega=\omega-m\Omega$, matches the slow magnetosonic wave
frequency. Significant wave energy/angular momentum absorption occurs
at both magnetic resonances, but with opposite signs, such that one of
them enhances the super-reflection while the other diminishes it. The
combined effect of the two magnetic resonances is to reduce the
super-reflection and the growth rate of the overstable p-modes. Our
calculations show that even a subthermal toroidal field (with the
magnetic pressure less than the gas pressure) may suppress the
overstability of hydrodynamic ($B=0$) p-modes. For accretion discs
with mixed (toroidal and vertical) magnetic fields, two additional
\Alfven resonances appear, where $\tomega$ matches the local \Alfven
wave frequency. The effect of these additional resonances is to further
reduce or diminish the growth rate of p-modes. Our results suggest
that in order for the non-axisymmetric p-modes to be a viable
candidate for the observed high-frequency QPOs, the disc magnetic
field must be appreciably subthermal, or other mode excitation
mechanisms are at work.

\end{abstract}


\section{Introduction}

Quasi-periodic oscillations (QPOs) in black-hole (BH) X-ray binaries
have been intensively studied since the launch of NASA's
{\it Rossi X-ray Timing Explorer} (Swank 1999). The observed QPOs can be
divided into two classes. The low-frequency QPOs (about 0.1--50~Hz) are
common, observable when the systems are in the hard state and the
steep power-law state (also called ``very high state'';
see Done et al.~2007).
They typically have high amplitudes and high coherence
($Q\go 10$), and can vary in frequency on short timescales (minutes). The
high-frequency QPOs (HFQPOs, 40--450~Hz), however, have only
been observed in the steep power-law state in seven BH X-ray binaries.
They typically have low amplitudes ($0.5-2\%$ rms at 2-60~keV) and
low coherence ($Q\sim 2-10$), with small variations in frequency
(less than $\sim 15\%$), and appear alone or in pairs with a
frequency ratio close to 2:3 (see Remillard \& McClintock 2006 for a review).
QPOs in X-ray emission have also been detected in the active galaxy
RE J1034+396 (Gierlinski et al.~2008; Middleton et al.~2009)
as well as in several Ultra-Luminous X-ray sources (ULXs) (e.g.,
Strohmayer \& Mushotzky 2009; Feng, Rao \& Kaaret 2010).
These are likely the ``supermassive'' or ``massive'' analogs of the QPOs
detected in Galactic BH X-ray binaries (see Middleton \& Done 2010).

It is widely accepted that HFQPOs are associated with the dynamics
of BH accretion flows near the Inner-most Stable Circular Orbit (ISCO).
Beyond this general consensus, the physical origin of HFQPOs remains unclear,
despite their well-defined phenomenology.
In recent years, a number of models/theories with various degrees of
sophistication have been proposed to explain HFQPOs
[see Sec.~1 of Lai \& Tsang 2009 (hereafter LT09) for a
review of most of the existing models].
Of particular interest to the present paper
is the relativistic diskoseismic oscillation
model, in which general relativity (GR) effects produce trapped
oscillation modes in the inner region of the disc (e.g., Okazaki et al.~1987;
Nowak \& Wagoner 1991; see Wagoner 1999 and Kato 2001 for reviews).
One can divide various diskoseismic modes according
to their vertical structures (see Fig.~1 of Fu \& Lai 2009):
The so-called p-modes (also called inertial-acoustic modes)
have zero node in their wavefunctions along the vertical direction,
while the g-modes (also known as inertial modes or inertial-gravity modes)
\footnote{Note that other than similar mathematical forms of dispersion relations, disk g-modes have no relation to stellar g-modes which are driven by buoyancy.}
have at least one vertical nodes\footnote{The so-called c-modes
(corrugation modes) also have at least one vertical node, but they have
low frequencies and are not relevant to HFQPOs.}.
Diskoseismic g-modes have received wide theoretical attentions because
they can be trapped by the GR effect without relying on special
inner disc boundary conditions
(Okazaki et al.~1987), and they may be
resonantly excited by global disc deformations (warp and eccentricity)
through nonlinear effects (Kato 2003b, 2008; Ferreira \& Ogilvie 2008).
Kato (2003a) and Li, Goodman \& Narayan (2003) (see
also Zhang \& Lai 2006 and Latter \& Balbus 2009)
showed that the non-axisymmetric
g-mode that contains a corotation resonance (CR, where the wave patten frequency
$\omega/m$ equals the background rotation rate $\Omega$;
here $\omega$ is the mode frequency and $m$ is the
azimuthal mode number) in the wave zone is
heavily damped (see Tsang \& Lai 2009a for a similar study for the
c-modes).  Thus the only non-axisymmetric g-modes of
interest are those trapped around the maximum of the ($\Omega+\kappa/m$) profile
(where $\kappa$ is the radial epicyclic frequency).
However, the frequencies of such modes, $\omega\simeq m\Omega(r_{\rm ISCO})$,
are too high (by a factor of 2-3) compared to the observed
values, given the measured mass and the
estimated spin parameter of the BH (Silbergleit \& Wagoner
2008; Tassev \& Bertschinger 2007). Another potential problem with the g-modes
is that their trapping zones (which arise from the maximum of
the $\kappa$ profile near the ISCO) can be destroyed by a weak (sub-thermal)
magnetic field (Fu \& Lai 2009). For example, the cavity of the $m=0$ g-mode
for a non-spinning BH vanishes when
$B_z^2/(4\pi\rho c_s^2)\go 0.1$ (where $B_z$ is the vertical magnetic
field, $\rho$ is the density and $c_s$ is the sound speed).
This elimination of the g-mode cavity is probably the reason
for the disappearance of g-modes in recent MHD simulations (O'Neill,
Reynolds \& Miller 2009). Moreover, several numerical simulations
suggest that g-modes may not survive vigorous disc turbulence
driven by magneto-rotational instabilities (MRI)
(e.g. Arras, Blaes \& Turner 2006; Reynolds \& Miller 2009).

The focus of this paper is on p-modes.  In an unmagnetized disc,
non-axisymmetric p-modes are trapped between the inner disc edge
(at the ISCO) and the inner Lindblad resonance (ILR) radius (where
$\tomega=\omega-m\Omega=-\kappa$).  So the existence of global p-modes
does require a partially reflective inner disc boundary (see
LT09). LT09 showed that such non-axisymmetric p-modes can be overstable
for two reasons: (1) Because the spiral density
waves inside the ILR carry negative
energies while those outside the outer Lindblad resonance (OLR) (where
$\tomega=\kappa$) carry positive energies, the leakage of the p-mode
energy when the wave tunnels through the corotation barrier (between
ILR and OLR) leads to mode growth.  (2) More importantly, p-modes
can become overstable due to wave absorption at the corotation
resonance (CR, where $\tomega=0$).  Near the BH, the radial
epicyclic frequency $\kappa$ reaches a maximum before decreasing to
zero at the ISCO. This causes a non-monotonic behavior in the fluid
vortensity, $\zeta=\kappa^2/(2\Sigma\Omega)$ (where $\Sigma$ is the
surface density of the disc), such that $d\zeta/dr>0$ inside the
radius where $\zeta$ peaks.  It can be shown that the sign of the
corotational wave absorption depends on the sign of $d\zeta/dr$ (Tsang
\& Lai 2008; LT09; see also Goldreich \& Tremaine 1979)
\footnote{This applies to barotropic discs, for which the
pressure is a function of the density $\rho$ only.
For a disc with radial stratification,
the radial entropy gradient affects wave absorption at the CR
(see Lovelace et al.~1999; Baruteau \& Masset 2008), and one can define
an effective vortensity, $\zeta_{\rm eff}=\zeta/S^{2/\Gamma}$ (where
$S=P/\Sigma^\Gamma$ and $\Gamma$ is the 2-dimensional adiabatic index),
that plays a similar role as $\zeta$ (see Tsang \& Lai 2009b).}.
Thus, p-modes with positive vortensity gradient at the corotation radius
can grow in amplitudes due to corotational wave absorption.
The resulting overstable non-axisymmetric p-modes were
suggested to offer a possible explanation for HFQPOs (LT09;
Tsang \& Lai 2009b).

Realistic BH accretion disks are expected to contain appreciable
magnetic fields as a result of the nonlinear development of the MRI
(e.g., Balbus \& Hawley 1998).
Starting from Hawley et al.~(1995), Brandenburg et al.~(1995) and
Stone et al.~(1996) in the 1990s,
the property of the MRI turbulence have been
studied using numerical simulations under different conditions and
setups (local shearing box with/without
vertical stratification, with/without explicit viscosity and resistivity,
different equations of state, global simulations; e.g.,
Miller \& Stone 2000; Hirose et al.~2009; Guan et al.~2009; Simon et al.~2009;
Davis et al.~2010; Fromang 2010; Longaretti \& Lesur 2010;
Sorathia et al.~2010).
Although the precise mechanism of the nonlinear saturation of the MRI
remains unclear (see Pessah \& Goodman 2009),
the simulations generally show that turbulent discs are
dominated by large-scale toroidal fields approaching (but less than)
equipartition, with somewhat smaller poloidal fields.
Therefore, it is important to understand how magnetic fields change the
characters of the CR
and the overstability of BH disc p-modes. These are
the main issues we will address in this paper.

We have recently shown (Fu \& Lai 2009),
based on local analysis of diskoseismic waves,
that the propagation diagram of p-modes, (particularly
the wave zone between inner disc edge (at $r_{\rm ISCO}$) and the ILR
are not significantly modified by disc magnetic fields.
Thus, we expect that the p-mode frequencies in a magnetic disc are not
much different from those in a zero-field disc.
However, the effect of magnetic field on the corotational wave
absorption is more complicated. As we will see, for a pure toroidal field,
the CR is split into two {\it magnetic slow resonances},
where the Doppler-shifted mode frequency $\tomega$ matches the
frequency of slow MHD waves propagating along the field line
(see Sakurai et al.~1991; Goossens et al.~1992; Terquem 2003). With a mixed
(poloidal and toroidal) magnetic field, two additional
{\it \Alfven resonances} appear, where $\tomega$ matches the
\Alfven frequency associated with the toroidal field
(Keppens et al.~2002).
Thus, in general, the original CR is split into
four resonances (the inner/outer magnetic slow resonances and the inner/outer
\Alfven resonances), which exhibit a rather rich and complex dynamical
behavior.

We note that Tagger and collaborators (Tagger \& Pellat 1999;
Varniere \& Tagger 2002; Tagger \& Varniere 2006) developed the theory of
accretion-ejection instability for discs containing large-scale
poloidal magnetic fields with strengths of order equipartition. Most
of their analysis were restricted to vertical fields threading an
infinitely thin disc with vacuum outside.  They showed that the
long-range behavior of the perturbed vacuum field outside the disc
greatly enhances the overstability of non-axisymmetric disc oscillation
modes. It is currently unclear whether the magnetic field configurations
adopted by Tagger et al. can be realized by MRI turbulence -- such field configuration may be produced by magnetic field advection from large radii (e.g. Lovelace et al. 2009). It is also
unclear if the accretion-ejection instability is robust when more
general disc field configurations are assumed.
By contrast, in this paper we focus on the effects of
magnetic fields {\it inside} the disc.
Since the p-modes we are interested in have no vertical structure
(i.e., $k_z=0$), our disc model
is essentially a cylinder, with the unperturbed disc properties
(e.g., the density $\rho$, sound speed $c_s$, rotation $\Omega$ and magnetic
fields $B_\phi$ and $B_z$) depending only on $r$ (the cylindrical radius).
While the magnetic field configurations considered in this paper
are prone to MRI, the p-modes
are not {\it directly} affected by the MRI because $k_z=0$.

Our paper is organized as follows. In Sec.~2, we introduce the equilibrium
state of the disc and general MHD equations with comments on the equation
structure. We present in Sec.~3 the effective potential of wave propagations
for different magnetic field configurations. Section 4 examines how
super-reflection across the corotation barrier is modified by the magnetic
field. We present numerical calculations of global p-modes of magnetized
BH accretion discs in Sec.~5 and conclude in Sec.~6.

\section{Setup and Basic Equations}

We consider a non-self-gravitating accretion disc, satisfying
the usual ideal MHD equations

\be
{\partial{\rho} \over \partial t}
+\nabla\cdot(\rho \bb{v})=0,
\label{eq:mhd1}
\ee
\be
{\partial{\bb{v}} \over \partial
t}+({\bb{v}}\cdot\nabla){\bb{v}}
=-\frac{1}{\rho}\nabla \Pi-{\nabla \Phi}
+\frac{1}{4\pi\rho}(\bb{B}\cdot\nabla)\bb{B},
\label{eq:mhd2}
\ee
\be
{{\partial {\bb{B}}} \over \partial t}=\nabla\times({\bb{v}}
\times{\bb{B}}),
\label{eq:mhd3}
\ee
where $\rho,\,P,\,\bb{v}$ are the fluid density, pressure and velocity,
$\Phi$ is the gravitational potential, and
\be
\Pi \equiv P+\frac{\bb{B}^2}{8\pi}
\label{eq:tp}
\ee
is the total pressure.
The magnetic field $\bb{B}$ also
satisfies the equation $\nabla\cdot\bb{B}=0$. We assume the flow obeys the barotropic
equation of state $P=P(\rho)$.

We adopt the cylindrical coordinates $(r, \phi, z)$ which
are centered on the central BH and have
the $z$-axis in the direction perpendicular to the disc plane.
The unperturbed background disc has a velocity field
$\bb{v}=r\Omega(r)\bb{\hat \phi}$,
and the magnetic field consists of both toroidal and vertical components $\bb{B}= B_\phi(r)\bb{\hat \phi}+B_z(r) \bb{\hat z}$. The gravitational acceleration in radial direction is defined as
\be
g=\frac{d\Phi}{dr}
\ee
so that $-\nabla\Phi=-g{\bb{\hat r}}$ and $g=r\Omega_{\rm K}^2>0$, where
$\Omega_{\rm K}$ is the angular frequency for a test mass (the Keplerian
frequency)\footnote{Here ``Keplerian'' does not necessarily mean the gravitational potential is Newtonian. See the end of Sec. 2.}.
Thus the radial force balance equation reads
\be
\rho g=\rho r\Omega^2-\frac{d\Pi}{dr}-\frac{B_{\phi}^2}{4\pi r}.
\label{eq:fb}
\ee

To probe the dynamical properties of the magnetized flow, we apply
linear perturbations to the ideal MHD Eqs.~(\ref{eq:mhd1})-(\ref{eq:mhd3})
by assuming the perturbation of any physical variable $f$ to have the form $\delta f \propto {\rm e}^{{\rm i}m\phi-{\rm i}\omega t}$ with $m$ being the azimuthal mode number and $\omega$ the
wave frequency. The background flow and magnetic field have no
$z$-dependance and we will assume that the perturbation also has no
$z$-dependance ($k_z=0$).
The resulting linearized perturbation equations clearly contain variables
$\delta \bb{v}$, $\delta \rho$, $\delta P$, $\delta \Pi$ and $\delta \bb{B}$.
To simply the algebra, we define a new variable
\be
\delta h=\frac{\delta \Pi}{\rho}=
\frac{\delta P}{\rho}+\frac{\bb{B}\cdot \delta \bb{B}}{4\pi\rho}.
\ee
Using $\Delta \bb{v}=\delta
\bb{v}+\bb{\xi}\cdot\nabla\bb{v}=d\bb{\xi}/dt
=-i\omega\bb{\xi}+(\bb{v}\cdot \nabla)\bb{\xi}$, we find that the
Eulerian perturbation $\delta \bb{v}$ is related to the Lagrangian
displacement vector $\bb{\xi}$ by $\delta
\bb{v}=-i\tomega\bb{\xi}-r\Omega'\xi_r \bb{\hat{\phi}}$ (prime denotes
radial derivative). Also, we have (for barotropic flow)
\be
\delta \rho=\delta P/c_s^2
\ee
with $c_{\rm s}=\sqrt{dP/d\rho}$ being the sound speed.
Therefore, we can express all the perturbation quantities in
terms of $\xi_r$ and $\delta h$, then further combine the perturbation
equations into two equations for these two variables:
\be
\frac{d\xi_r}{dr}=A_{11}\xi_{r}+A_{12}\delta h,
\label{eq:ode1}
\ee
\be
\frac{d\delta h}{dr}=A_{21}\xi_r+A_{22}\delta h,
\label{eq:ode2}
\ee
where
\be
A_{11}=\frac{r\tomega^2\left[(\omega_{A\phi}^2-\Omega^2)\tomega^2
+\omega_{A\phi}^2\omega^2\right]}{(c_s^2+v_A^2)(\tomega^2-m^2\omega_{A\phi}^2)
(\tomega^2-\omega_s^2)}
+\frac{g\tomega^2}{(c_s^2+v_A^2)(\tomega^2-\omega_s^2)}
-\frac{\tomega^2+2m\tomega\Omega+m^2\omega_{A\phi}^2}{r(\tomega^2-m^2\omega_{A\phi}^2)},
\label{eq:a11}
\ee
\be
A_{12}=-\frac{\tomega^4}{(c_s^2+v_A^2)(\tomega^2-m^2\omega_{A\phi}^2)
(\tomega^2-\omega_s^2)}+\frac{m^2}{r^2(\tomega^2-m^2\omega_{A\phi}^2)},
\label{eq:a12}
\ee
\[
A_{21}=\tomega^2-m^2\omega_{A\phi}^2-\frac{4(m\omega_{A\phi}^2+\tomega\Omega)^2}
{\tomega^2-m^2\omega_{A\phi}^2}+r\frac{d}{dr}(\omega_{A\phi}^2-\Omega^2)
+(\omega_{A\phi}^2-\Omega^2)\frac{r}{\rho}\frac{d\rho}{dr}
+\frac{g}{\rho}\frac{d\rho}{dr}
\]
\be
~~~~~~~+\frac{1}{(c_s^2+v_A^2)(\tomega^2-m^2\omega_{A\phi}^2)(\tomega^2-\omega_s^2)}
\left\{r\left[(\omega_{A\phi}^2-\Omega^2)\tomega^2+\omega_{A\phi}^2\omega^2\right]
+g(\tomega^2-m^2\omega_{A\phi}^2)\right\}^2,
\label{eq:a21}
\ee
\be
A_{22}=-\frac{r\tomega^2\left[(\omega_{A\phi}^2-\Omega^2)\tomega^2
+\omega_{A\phi}^2\omega^2\right]}{(c_s^2+v_A^2)(\tomega^2-m^2\omega_{A\phi}^2)
(\tomega^2-\omega_s^2)}
-\frac{g\tomega^2}{(c_s^2+v_A^2)(\tomega^2-\omega_s^2)}
+\frac{2m(m\omega_{A\phi}^2+\tomega\Omega)}{r(\tomega^2-m^2\omega_{A\phi}^2)}
-\frac{1}{\rho}\frac{d\rho}{dr}.
\label{eq:a22}
\ee
In the above expressions,
\be
\tomega=\omega-m\Omega
\ee
is the wave frequency in the co-rotating frame,
\be
v_A={|\bb{B}|\over \sqrt{4\pi \rho}}
\ee
is the \Alfven velocity,
\be
\omega_{A\phi}=\frac{v_{A\phi}}{r}={B_{\phi}\over r\sqrt{4\pi \rho}}
\ee
is the toroidal \Alfven frequency, and
\be
\omega_s=\frac{c_s}{\sqrt{c_s^2+v_A^2}}m\omega_{A\phi}=
k_{\phi}\frac{c_s v_{A\phi}}{\sqrt{c_s^2+v_A^2}},
\label{eq:sf}
\ee
is the slow magnetosonic wave frequency for $\bb{k}=(m/r)\bb{\hat{\phi}}$.

Equations (\ref{eq:ode1})-(\ref{eq:ode2}) describe the linear
perturbations (without vertical wavenumber) in a compressible 2D flow
threaded by a mixture of toroidal and vertical magnetic fields. In the
hydrodynamic disc limit ($B\rightarrow 0$), they reduce to the related
equations in Goldreich \& Tremaine (1979) with no external forcing
potential [see also Tsang \& Lai 2008 (TL08 hereafter) and LT09].
They also agree with system (14) in Blokland et al.~(2005)
\footnote
{There is a typo in their Eq.~(20), where in the last line, the term
  $(B_{\theta}^2\tomega+Fv_{\theta})$ should be
  $(B_{\theta}^2\tomega+Fv_{\theta})^2$.}  in the $k_z \rightarrow 0$
and barotropic limit, although their equations are cast on slightly
different variables ($\delta \Pi$ instead of $\delta h$). The
Hameiri-Bondeson-Iacono-Bhattacharjee (HBIB) type of equations
presented in Blokland et al.~(2005) have been derived before by
Hameiri (1981) and Bondeson et al.~(1987) [see also Mikhailovskii et
  al.~(2009); Goossens et al.~(1992) (hereafter GHS92) and reference
  therein] for a magnetized rotating flow in the absence of external
gravity. The equations including gravity were obtained by Keppens et
al.~(2002) (hereafter KCP02) using the same Frieman-Rotenberg
technique (Frieman \& Rotenberg 1960).  System (5) in KCP02 is by far
the most general formulation (with finite $B_{\phi}$, $B_z$, $m$,
$k_z$, $v_{\phi}$, $v_z$ and adiabatic equation of state) that in
principle governs all the MHD waves and instabilities for a
compressible MHD fluid with an equilibrium flow in cylindrical
geometry. In this paper, we focus on the effects of magnetic fields on
the corotational instability of the inertial-acoustic modes (p-modes)
in accretion discs, thus Eqs.~(\ref{eq:ode1})-(\ref{eq:ode2}) are our
main working equations.  Recently, Yu \& Li (2009) considered a
similar system (with finite $k_z$ and pure toroidal B field) to study
the Rossby wave instability and they obtained equations with similar
structure as Eqs.~(\ref{eq:ode1})-(\ref{eq:ode2}).  Our equations in
the pure toroidal B field limit ($B_z \rightarrow 0$) are the same as
theirs in the $k_z\rightarrow 0$ limit.
Also, the disc system that we study here are similar to the one considered by Terquem (2003), where instead
of solving for global modes the author looked into the interaction
between the disc and a planet and the consequences on planet
migrations.

Evidently, the coefficients in Eqs.~(\ref{eq:a11})-(\ref{eq:a22}) are singular when
\be
\tomega^2=m^2\omega_{A\phi}^2,
\label{eq:afr}
\ee
and
\be
\tomega^2=\omega_s^2.
\label{eq:slr}
\ee
In more general situations (with $k_z \neq 0$), $m\omega_{A\phi}$ in
Eq.~(\ref{eq:afr}) generalizes to
($\bb{k}\cdot\bb{B})/\sqrt{4\pi\rho}$, which is the \Alfven frequency,
and the slow magnetosonic frequency $\omega_s$ generalizes to
$\sqrt{c_s^2/(c_s^2+v_A^2)} (\bb{k}\cdot
\bb{B})/\sqrt{4\pi\rho}$. Following Sakurai et al. (1991) and KCP02,
we shall call (\ref{eq:afr}) the {\it \Alfven Resonance} (AR) and
(\ref{eq:slr}) the {\it Magnetic slow Resonance} (MR). Note that these singularities/resonances have also been studied in the specific context of disc-planet interactions. Terquem (2003) considered a 2D disk with a pure toroidal magnetic field and found that a subthermal toroidal B-field is capable of stopping the inward planetary migration as long as the $B_{\phi}(r)$ gradient is negative enough. This analytic finding was later confirmed by 2D MHD simulation results (Fromang et al. 2005). On the other hand, Muto et al. (2008), by employing shearing sheet approximation, performed 3D (i.e., $k_z \neq 0$) calculations for a disk with a pure poloidal magnetic field and showed that type I planetary migration could be outward if the B-filed is strong enough (superthermal).

In the special case of $k_z=B_z=0$, Eq.~(\ref{eq:afr}) is no longer a real singularity
of the differential equations. In this case, one can show, through some subtle
mathematical manipulations, that the terms
$\tomega^2-m^2\omega_{A\phi}^2$ in $A_{11}$, $A_{12}$, $A_{21}$ and $A_{22}$
all get canceled. These four coefficients then reduce to
\be
A_{11}=\frac{v_{A\phi}^2}{c_s^2+v_{A\phi}^2}
\frac{1}{\tomega^2-\omega_s^2}
\left[\frac{\tomega^2}{r}-\frac{m^2c_s^2}{r^3}
-\frac{c_s^2}{v_{A\phi}^2}\frac{\tomega^2+2m\tomega\Omega}{r}
-\frac{\tomega^2}{v_{A\phi}^2}(r\Omega^2-g)\right],
\label{eq:A11}\ee
\be
A_{12}=-\frac{1}{v_{A\phi}^2+c_s^2}
\frac{\tomega^2-m^2c_s^2/r^2}{\tomega^2-\omega_s^2},
\ee
\[
A_{21}=\tomega^2-m^2\omega_{A\phi}^2-\kappa^2
-\frac{d\ln (\rho r/B_{\phi}^2)}{d\ln r}
\frac{v_{A\phi}^2}{c_s^2+v_{A\phi}^2}(\Omega^2-g/r-2c_s^2/r^2)
\]
\be
~~~~~~~-\frac{1}{\tomega^2-\omega_s^2}
\Big\{4\Omega^2\omega_s^2+\left[\frac{mv_{A\phi}^2}{c_s^2+v_{A\phi}^2}
(\Omega^2-g/r+2c_s^2/r^2)\right]^2
+4m\tomega\Omega\frac{v_{A\phi}^2}{c_s^2+v_{A\phi}^2}
(\Omega^2-g/r+2c_s^2/r^2)\Big\},
\ee
\be
A_{22}=\frac{m}{r}\frac{c_s^2}{c_s^2+v_{A\phi}^2}
\frac{1}{\tomega^2-\omega_s^2}\left[
2\tomega\Omega+\frac{mv_{A\phi}^2}{c_s^2+v_{A\phi}^2}
(\Omega^2-g/r+2c_s^2/r^2)\right]
-\frac{d\ln \rho}{d\ln r}+\frac{1}{c_s^2+v_{A\phi}^2}
(r\Omega^2-g-2v_{A\phi}^2/r),
\label{eq:A22}\ee
where the radial epicyclic frequency $\kappa$ is given by
\be
\kappa=\left[\frac{2\Omega}{r}\frac{d}{dr}(r^2\Omega)\right]^{1/2}.
\ee
This peculiar disappearance of \Alfven singularity can also be verified from a different
perspective. GHS92 derived the jump conditions for $\xi_r$ and $\delta \Pi$
(called $P'$ in their paper) across the \Alfven resonance
\[
[\xi_r] \propto g_{B}=mB_z/r-k_zB_{\phi},
\]
\[
[\delta \Pi]\propto B_z
\]
(see their Eqs.~[42]-[43]). We see that when $k_z=B_z=0$,
both $\xi_r$ and $\delta \Pi$ are continuous across the
\Alfven resonance, thus the \Alfven  singularity disappears
\footnote{Although the analysis in GHS92 does not consider external
gravity, one can show that the effect of gravity can be easily included through
replacing the term $\rho v_{\phi}^2/r$ in their Eq.~(5) by
$\rho v_{\phi}^2/r+\rho g$. This modification only changes the relations between
the background variables ($B_{\phi}$, $v_{\phi}$, $c_s$, etc.) The jump conditions for
$\xi_r$ and $P'$ would remain the same.}. In a non-magnetic disc,
Eqs.~(\ref{eq:afr})-(\ref{eq:slr}) reduce to the corotation singularity (TL08).

In Sec. 3 and Sec. 4, we will simply employ Newtonian gravitational potential, which has free-particle orbital frequency $\Omega_{\rm K} \propto r^{-3/2}$ and epicyclic frequency $\kappa=\Omega_{\rm K}$. This way a dimension-free analysis can be easily made without missing the essential physics. However, in Sec. 5, in order to make a direct comparison with results in LT09, we will use the
Paczynski-Wiita pseudo-Newtonian potential (Paczynski \& Wiita 1980)
\be
\Phi=-\frac{GM}{r-r_{\rm s}},
\label{eq:pwp}
\ee
where $M$ is the central BH mass and $r_{\rm s}=2GM/c^2$ is the Schwarzschild
radius. The corresponding free-particle orbital frequency and radial epicyclic frequency are
\be
\Omega_{\rm K}=\left(\frac{1}{r}\frac{d\Phi}{dr}\right)^{1/2}=
\sqrt{\frac{GM}{r}}\frac{1}{r-r_{\rm s}},
\label{eq:oef}
\ee
\be
\kappa=\Omega_{\rm K}\sqrt{\frac{r-3r_{\rm s}}{r-r_{\rm s}}}.
\label{eq:epicyclic}
\ee
Note that $\kappa^2$ peaks at $r=(2+\sqrt{3})r_{\rm s}$ and decreases
to zero at $r=3r_{\rm s}$.
Throughout this paper, we will consider thin discs with
the sound speed $c_s\ll r\Omega_{\rm K}$ and
magnetic field satisfying $B^2/(4\pi\rho)\ll (r\Omega_{\rm K})^2$, so that
$\Omega\simeq\Omega_{\rm K}$.

\section{Effective potential and Wave Propagation diagram}

We can combine Eqs.~(\ref{eq:ode1})-(\ref{eq:ode2}) into a single second-order
differential equation on $\delta h$
\be
\frac{d^2}{dr^2}\delta h+C_1(r)\frac{d}{dr}\delta h+C_0(r)\delta h=0,
\label{eq:sse}
\ee
where
\be
C_1(r)=-A_{11}-A_{22}-\frac{1}{A_{21}}\frac{dA_{21}}{dr},
\label{eq:C1}\ee
\be
C_0(r)=A_{11}A_{22}-A_{12}A_{21}-\frac{dA_{22}}{dr}+\frac{A_{22}}{A_{21}}
\frac{dA_{21}}{dr}.
\label{eq:C0}\ee
We introduce a new variable
\be
\eta=\mbox{exp}\left[\int^r\frac{1}{2}C_1(r)dr\right]\delta h,
\label{eq:eta}
\ee
so that Eq.~(\ref{eq:sse}) can be rewritten as a wave equation for $\eta$:
\be
\frac{d^2\eta}{dr^2}-V(r)\eta=0,
\label{eq:waveeq}\ee
where
\be
V(r)=\frac{1}{4}C_1(r)^2+\frac{1}{2}\frac{dC_1(r)}{dr}
-C_0(r)
\label{eq:effp}
\ee
is the effective potential for wave propagation.
Considering a local plane wave solution
\be
\eta \propto \mbox{exp}\left[{\rm i}\int^rk_r(s) ds\right],
\ee
Eq.~(\ref{eq:waveeq}) then yields
\be
k_r^2=-V(r).
\label{eq:kr}
\ee
Formally, wave can propagate only in the region with $V(r) < 0$.
Our derivation above is quite general and can be applied to complicated systems
as the details have all been captured in the four coefficients $A_{ij}$.
Figures \ref{fig:poten1}-\ref{fig:poten3}
show the effective potential for various situations
discussed below.

\subsection{Hydrodynamical discs}

\begin{figure}
\begin{center}
\includegraphics[width=0.7\textwidth]{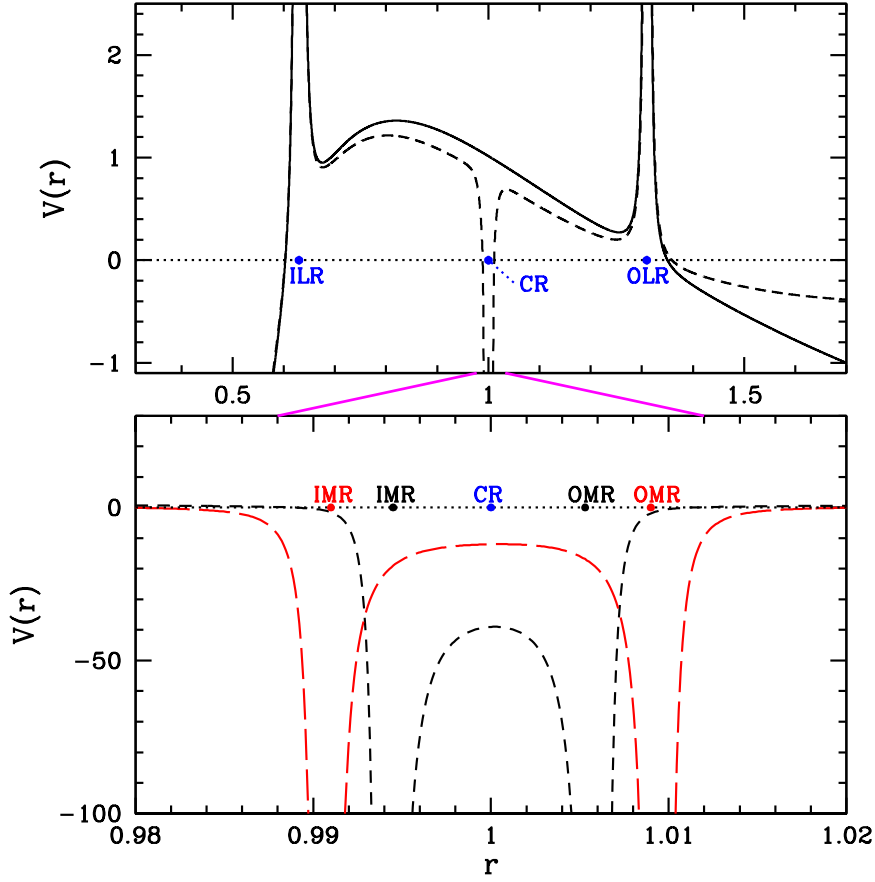}
\caption{Wave propagation diagram in accretion discs when the
  background fluid vortensity gradient (evaluated at the corotation
  resonance radius) $(d\zeta/dr)_{\rm c}=0$.  In the upper panel, the
  solid line shows the effective potential $V(r)$ as a function of $r$
  for $B=0$ discs, and the dashed line shows $V(r)$ for discs with
  finite toroidal magnetic fields. Waves can propagate only in the
  region where $V(r)<0$. The labels ILR, OLR and CR denote the inner
  Lindblad resonance, outer Lindblad resonance and corotation
  resonance, respectively. The lower panel shows the blow-up of the
  region near the CR (located at $r_{\rm c}=1$), and the long-dashed line
  shows the case with a stronger $B_\phi$ than the short-dashed line.
  The labels IMR and OMR denote the inner magnetic (slow) resonance
  and outer magnetic resonance, respectively.  Note that the vertical
  scales in the upper and bottom panels differ by a large factor. A
  Newtonian potential is used in calculating $V(r)$.  (A color version
  of this figure is available in the online journal.)}
\label{fig:poten1}
\end{center}
\end{figure}

\begin{figure}
\begin{center}
\includegraphics[width=0.7\textwidth]{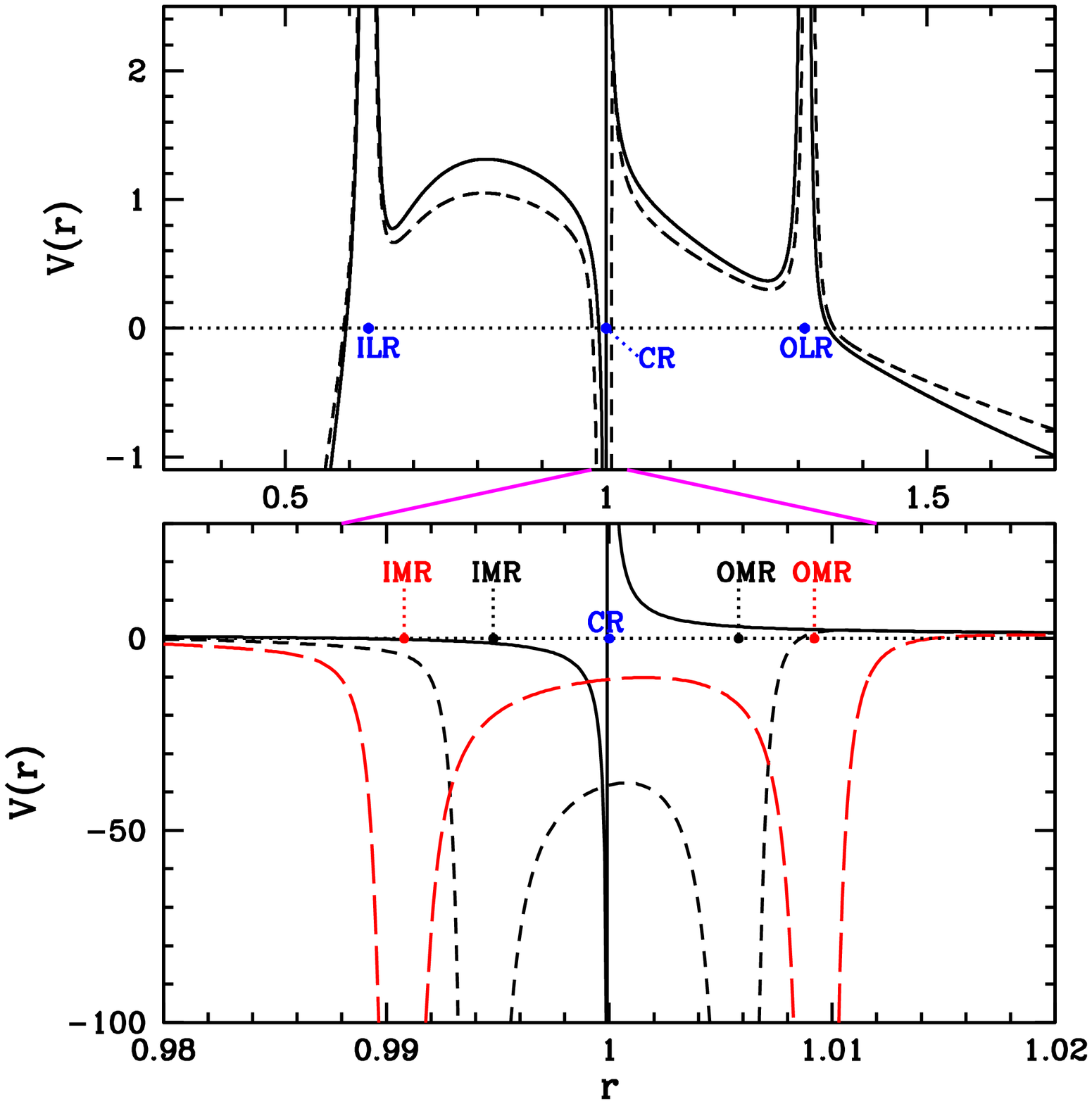}
\caption{Same as Fig.~\ref{fig:poten1}, except for the case
$(d\zeta/dr)_{\rm c}<0$.}
\label{fig:poten2}
\end{center}
\end{figure}

\begin{figure}
\begin{center}
\includegraphics[width=0.7\textwidth]{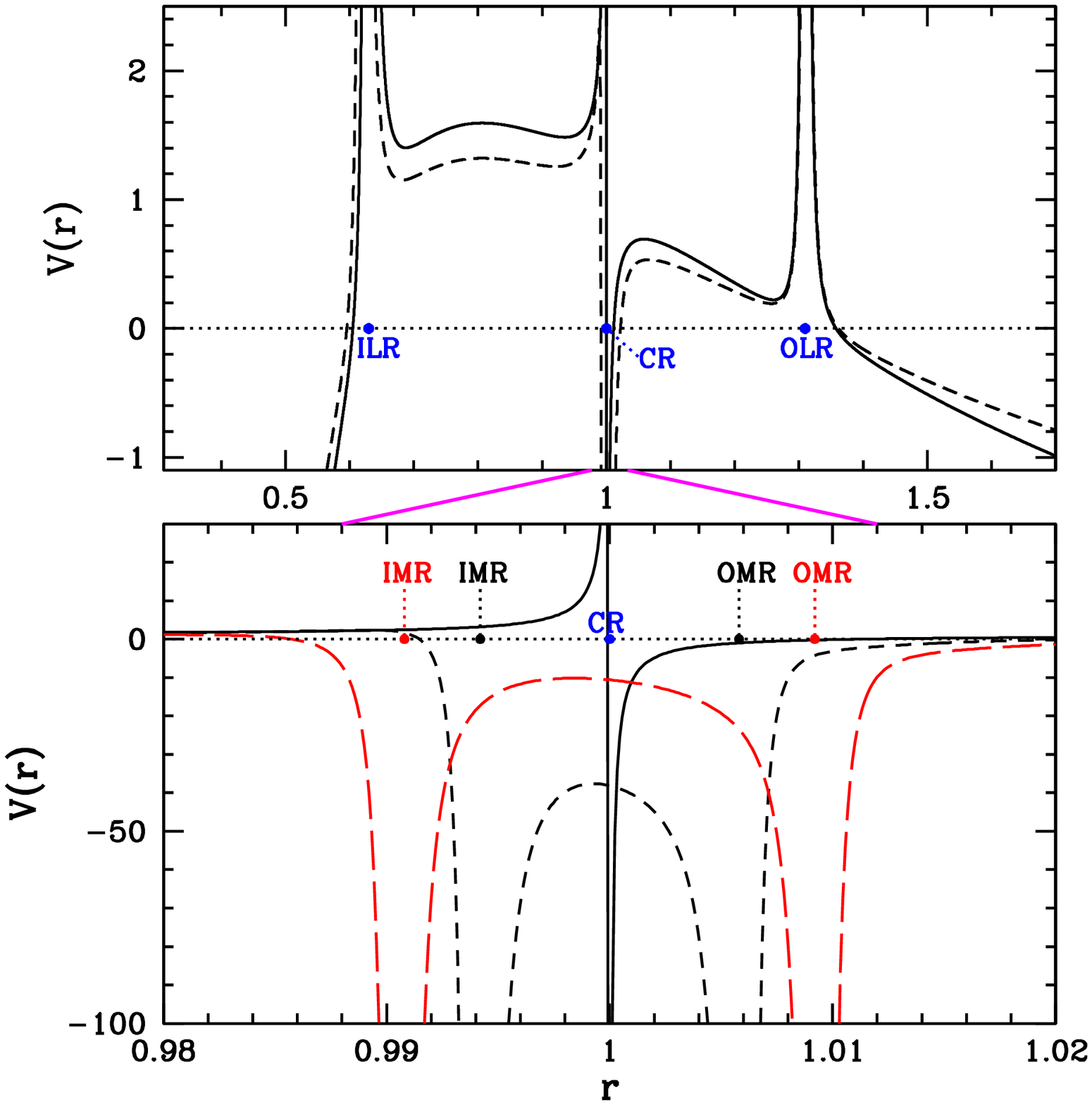}
\caption{Same as Fig.~\ref{fig:poten1}, except for the case
$(d\zeta/dr)_{\rm c}>0$.}
\label{fig:poten3}
\end{center}
\end{figure}

For a non-magnetic disc, the four coefficients $A_{ij}$ simplify to
\be
A_{11}=-\frac{d\ln (r\rho)}{dr}-\frac{2m\Omega}{r\tomega},
\ee
\be
A_{12}=-\frac{1}{c_s^2}+\frac{m^2}{r^2\tomega^2},
\ee
\be
A_{21}=-D,
\ee
\be
A_{22}=\frac{2m\Omega}{r\tomega},
\ee
where $D=\kappa^2-\tomega^2$.
Thus $C_0(r)$ and $C_1(r)$ now read
\be
C_0(r)=-\frac{D}{c_s^2}-\frac{m^2}{r^2}-\frac{2m\Omega}{r\tomega}
\left(\frac{d}{dr}\ln \frac{\Omega \rho}{D}\right),
\ee
\be
C_1(r)=-\frac{d}{dr}\ln \frac{D}{r\rho}.
\ee
The effective potential then takes the following form
\be
V(r)=\frac{D}{c_s^2}+\frac{m^2}{r^2}+
\frac{2m\Omega}{r\tomega}\left(\frac{d}{dr}\ln \frac{\Omega \rho}{D}\right)
+S^{1/2}\frac{d^2}{dr^2}S^{-1/2},
\label{eq:hdpoten}
\ee
where $S=D/(r\rho)$.
The effective potential $V(r)$ is singular
at both the corotation resonance (CR), where $\tomega=0$, and the Lindblad
resonances (LRs), where $D=0$
(see Figs.~\ref{fig:poten1}-\ref{fig:poten3}).
However, the singularities at the LRs
are spurious singularities of the wave equation (i.e. they can be removed by rewriting the equation in alternative variables [e.g., Narayan et al.~1987; TL08])
and only the CR is a non-removable singularity. The
behavior of $V(r)$ around the CR depends on $d\zeta/dr$ (evaluated
at the corotation radius $r_{\rm c}$), the
gradient of the background fluid vortensity, $\zeta$, defined by
\be
\zeta=\frac{\kappa^2}{2\Omega\rho}.
\label{eq:zeta}
\ee
For $d\zeta/dr<0$ a narrow ``Rossby wave zone'' lies just inside $r_{\rm c}$,
while for $d\zeta/dr>0$ the Rossby zone lies just outside $r_{\rm c}$;
in the special case of $d\zeta/dr=0$, the corotation singularity disappears (see TL08).

\begin{figure}
\begin{center}
\includegraphics[width=0.7\textwidth]{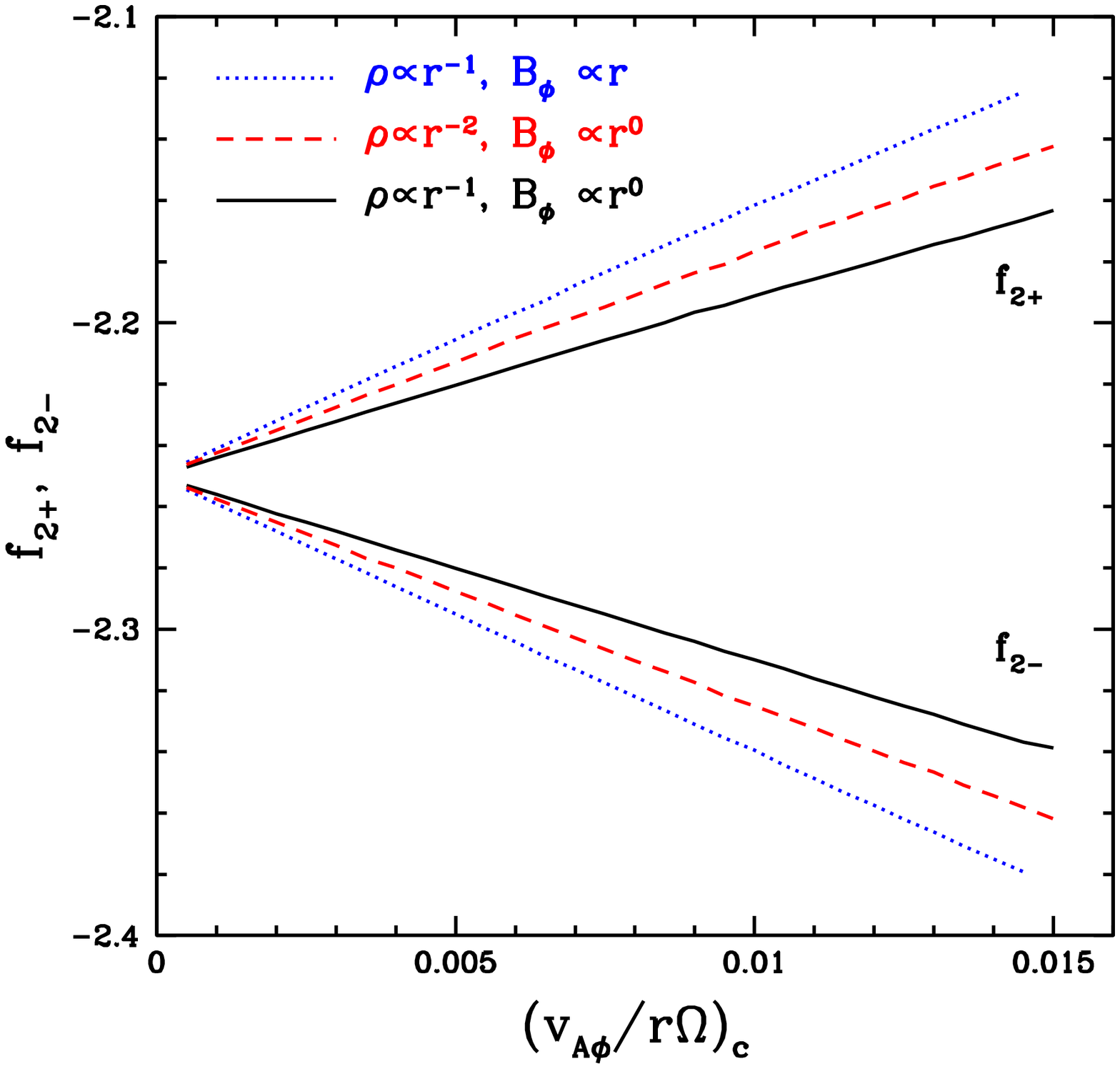}
\caption{Numerical values of $f_{2+}$ and $f_{2-}$
[see Eq.~(\ref{eq:coeq})] as a function of the dimensionless
ratio $v_{A\phi}/(r\Omega)$ (evaluated at $r=r_{\rm c}$), which specifies
the toroidal magnetic field strength in the disc.
Dimensionless units are adopted, where $G=M=r_{\rm c}=1$, $m=2$,
and $c_s/(r\Omega)\ll 1$. Different line types represent different disc density and magnetic field profiles.}
\label{fig:coef}
\end{center}
\end{figure}

\subsection{Discs with pure toroidal magnetic fields}

As mentioned in Sec.~2, for a disc with a pure toroidal magnetic field,
the CR splits into two magnetic slow resonances (MRs). This can be seen clearly
from the wave propagation diagram shown in
Figs.~\ref{fig:poten1}-\ref{fig:poten3}.
For $B=0$, the effective potentials reduce to those shown in TL08.
In the upper panels of Figs.~\ref{fig:poten1}-\ref{fig:poten3},
we see that for small magnetic fields
(with $\omega_{A\phi} \ll \Omega$), the effective potential profile is not
affected much by the B-field except near the corotation. The blow-ups in
the bottom panels show that the toroidal magnetic field splits the CR
into two MRs, the inner magnetic resonance (IMR) and
outer magnetic resonance (OMR), one on each side of the CR.
A larger toroidal field (see the long-dashed lines) gives a wider
radial separation between the MRs and a shallower potential
well around $r_{\rm c}$. Note that independent of the sign of $(d\zeta/dr)_{\rm c}$,
there always exists an effective wave zone centered around corotation
point and bounded by the two MRs.

Far away from the resonances (the regions on the left and right sides
of the upper panels of
Figs.~\ref{fig:poten1}-\ref{fig:poten3}),
the WKB dispersion relation (\ref{eq:kr}) reduces to
(see Terquem 2003; Fu \& Lai 2009)
\be
\tomega^2=\kappa^2+k_r^2(c_s^2+v_{A\phi}^2).
\label{eq:disp}
\ee
This describes fast magnetosonic waves modified by the disc rotation. Thus the spiral density waves
(inertial-acoustic waves) inside the ILR and outside the OLR are
inertial-fast magnetosonic waves.

The behavior of the effective potential around the MRs is more difficult to
analyze. The IMR and OMR are separated from the CR by the distance
\be
r_{\rm mr}-r_{\rm c}\simeq \pm \left({2\omega_s\over 3m\Omega}\right)r_{\rm c},
\ee
(the upper sign is for the OMR and the lower sign for the IMR).
Because of the complexity of the equations involved, no simple
analytical expression for $V(r)$ can be derived, even in the asymptotic limit
(e.g., $\tomega\rightarrow\pm\omega_s$ and $v_{A\phi}/r\Omega\ll 1$).
Nevertheless, a careful examination of Eqs.~(\ref{eq:A11})-(\ref{eq:A22}),
(\ref{eq:C1})-(\ref{eq:C0}), (\ref{eq:effp})
and numerical calculation of $V(r)$
show that in the neighborhood of the MRs, the effective potential can be
written in the following form:
\be
V(r)=f_{0}+\frac{f_{1}}{\tomega^2-\omega_s^2}
+\frac{f_{2+}}{(\tomega-\omega_s)^2}
+\frac{f_{2-}}{(\tomega+\omega_s)^2},
\label{eq:coeq}
\ee
where $f_0,\,f_1,\,f_{2\pm}$ vary slowly around the CR/MRs region.
Thus, the effective potential is dominated by second-order
singularities at the MRs, i.e., $V(r)\propto (r-r_{\rm mr})^{-2}$
(see Figs.~\ref{fig:poten1}-\ref{fig:poten3}).
The coefficients $f_{2\pm}$ can be calculated numerically, and
some examples of how $f_{2+}$ and $f_{2-}$ depend the magnetic field
strength are shown in Fig.~\ref{fig:coef}. We see that in
dimensionless units (with $G=M=r_{\rm c}=1$, $m=2$ and $c_s\ll r\Omega$,
$v_{A\phi}\ll r\Omega$), both $f_{2+}$ and $f_{2-}$ are negative and
approximately equal to $-2$. Thus, near one of the MRs (e.g., the OMR),
we have $k_r^2=-V(r)\sim
2/(\tomega-\omega_s)^2\sim 0.2/(r-r_{\rm mr})^2$, or
$k_r\sim 0.4/|r-r_{\rm mr}|$. Since $k_r^2$ is smaller than $|dk_r/dr|$,
the WKB analysis is not really valid near the MRs. On the other hand,
close to the CR radius ($r=r_{\rm c}$), we have $V(r)\sim -4/\omega_s^{2}$,
which implies $k_r\sim 2\omega_s^{-1}$. Since the separation between
the IMR and OMR is $2\omega_s/3\Omega$ (for $m=2$), we find that the WKB
phase variation across the CR region (but avoiding the MRs) is of order unity.
Therefore, despite the deep effective potential at the MRs, physically there
is not much a wave zone around the MRs/CR region. This feature is
borne out in our numerical calculations of global wave modes
presented in Sec.~5.

It is worth noting with the second-order singularities
associated with the MRs, the $B\rightarrow 0$ limit is approached
in a non-trivial way. From Fig.~\ref{fig:coef}, we see that
as the magnetic field decreases, $f_{2+}$ and $f_{2-}$ reach
the same values ($\simeq -2.25$). Thus the singularities
associated with $f_{2+}/(\tomega-\omega_s)^2$ and
$f_{2-}/(\tomega+\omega_s)^2$ simply become $1/\tomega^2$ as $B\rightarrow 0$.
On the other hand, $f_{1}$ can be written as $f_1=f_{10}\tomega^0
+f_{11}\tomega+f_{12}\tomega^2+\cdots$, with $f_{1i}$ being near constant
around the MRs. As $B\rightarrow 0$, we have $f_{10}+f_{2+}+f_{2-}=0$,
and the second-order singularity term ($\propto 1/\tomega^2$) in $V(r)$
disappears. What is left is the term $f_{11}/\tomega$, a first-order
singularity associated with the CR in a non-magnetic disc. Evidently,
as $B\rightarrow 0$, $f_{11}$ is proportional to the vortensity gradient,
$d\zeta/dr$.

\subsection{Discs with pure vertical magnetic fields}

Before examining the case of mixed magnetic fields, it is useful
to consider discs with pure vertical fields. Here
we assume that $B_z$ is constant (independent of $r$)
to simplify the analysis. The coefficients $A_{ij}$ of the wave
equations (\ref{eq:ode1})-(\ref{eq:ode2}) are
\be
A_{11}=-\frac{c_s^2}{c_s^2+v_A^2}\frac{d\ln \rho}{dr}
-\frac{2m\Omega}{r\tomega}-\frac{1}{r},
\ee
\be
A_{12}=-\frac{1}{c_s^2+v_A^2}+\frac{m^2}{r^2\tomega^2},
\ee
\be
A_{21}=\tomega^2-\kappa^2-\frac{c_s^2v_A^2}{c_s^2+v_A^2}\left(\frac{d\ln \rho}{dr}\right)^2,
\ee
\be
A_{22}=-\frac{v_A^2}{c_s^2+v_A^2}\frac{d\ln \rho}{dr}
+\frac{2m\Omega}{r\tomega}.
\ee
We then have
\be
C_1=-\frac{d}{dr}\ln \left(\frac{A_{21}}{{r\rho}}\right),
\ee
\[
C_0=-\frac{D}{c_s^2+v_A^2}-\frac{m^2}{r^2}
-\frac{2m\Omega}{r\tomega}\left(\frac{d}{dr}\ln
\frac{\Omega\rho^{\lambda}}{A_{21}}\right)\nonumber
\]
\be
\quad
+\frac{m^2}{r^2\tomega^2}\frac{c_s^2v_A^2}{c_s^2+v_A^2}\left(\frac{d\ln \rho}
{dr}\right)^2-\frac{v_A^2}{c_s^2+v_A^2}\left(\frac{d\ln \rho}{dr}\right)
\left(\frac{d\ln (A_{21}/r)}{dr}\right)
+\frac{d}{dr}\left(\frac{v_A^2}{c_s^2+v_A^2}\frac{d\ln \rho}{dr}\right),
\ee
where
\be
\lambda=(c_s^2-v_A^2)/(c_s^2+v_A^2)
\ee
and $D=\kappa^2-\tomega^2$.
The resulting effective potential reads
\[
V(r)=\frac{D}{c_s^2+v_A^2}+\frac{m^2}{r^2}
+\frac{2m\Omega}{r\tomega}\left(\frac{d}{dr}\ln \frac{\Omega\rho^{\lambda}}{A_{21}}\right)
+S^{1/2}\frac{d^2}{dr^2}S^{-1/2}
\]
\be
~~~~~~~~-\frac{m^2}{r^2\tomega^2}\frac{c_s^2v_A^2}{c_s^2+v_A^2}\left(\frac{d\ln \rho}
{dr}\right)^2+\frac{v_A^2}{c_s^2+v_A^2}\left(\frac{d\ln \rho}{dr}\right)
\left(\frac{d\ln (A_{21}/r)}{dr}\right)
-\frac{d}{dr}\left(\frac{v_A^2}{c_s^2+v_A^2}\frac{d\ln \rho}{dr}\right),
\ee
where $S=A_{21}/\left(r\rho\right)$.
Compared with the $V(r)$ in the $B=0$ case, we see that a constant
vertical field does not split the CR. However, in contrast to the first-order corotation singularity ($1/\tomega$) in the $B=0$ case, the CR in discs with $B_z \neq 0$ manifests mainly as a second-order singularity ($\sim 1/\tomega^2$).

\subsection{Discs with mixed magnetic fields}

\begin{figure}
\begin{center}
\includegraphics[width=0.7\textwidth]{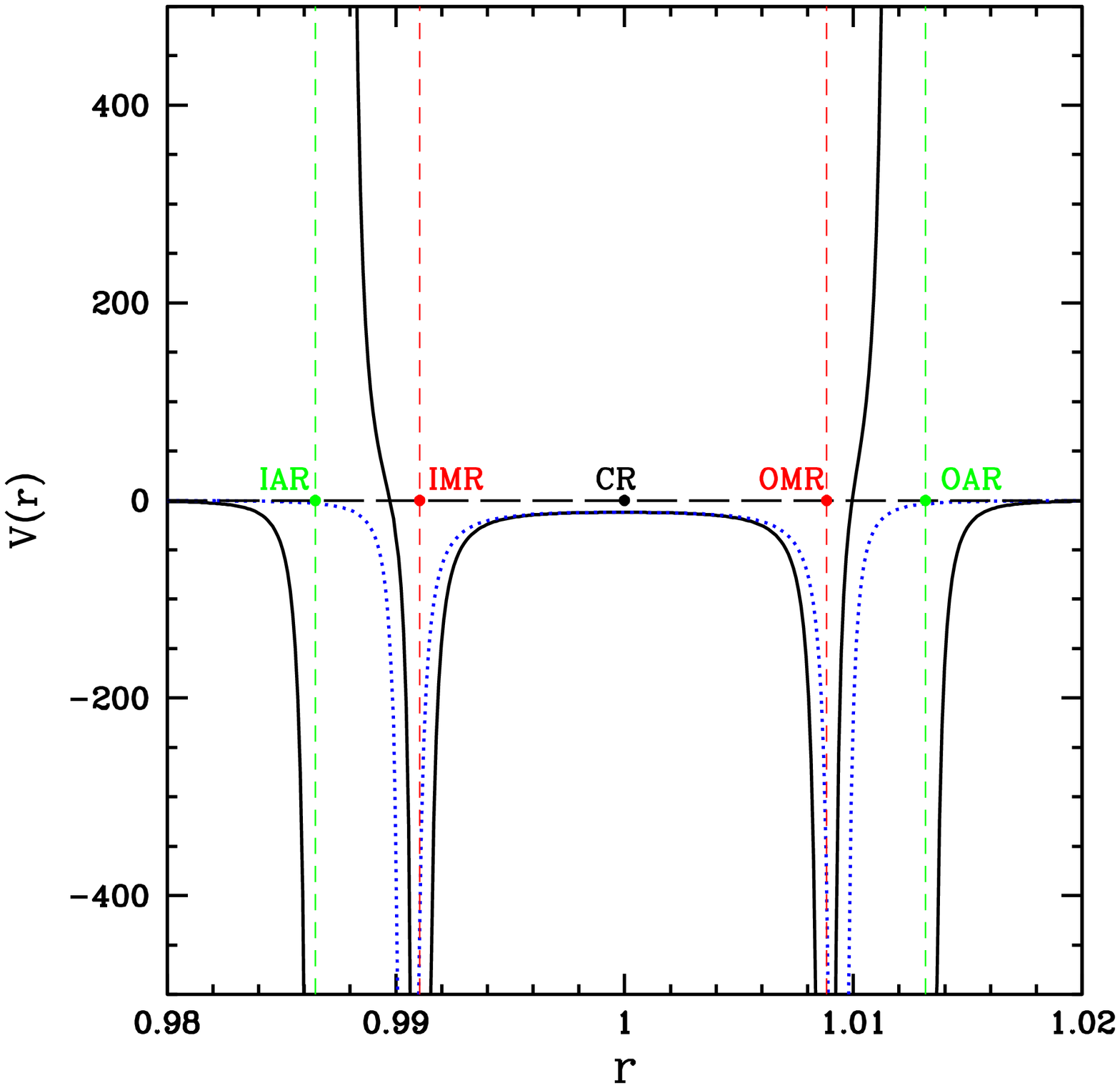}
\caption{Effective potential around the corotation/magnetic resonance
  region for a disc with mixed magnetic fields (solid line). The
  dotted line shows $V(r)$ for the case with a pure toroidal field.
  The corotation resonance (CR), inner/outer magnetic (slow)
  resonances (IMR and OMR) and inner/outer \Alfven resonances (IAR and
  OAR) are indicated.  Note that the MRs represent second-order
  singularities in $V(r)$, while the ARs (which exist only for the
  mixed field case) are first-order singularities.}
\label{fig:potenm}
\end{center}
\end{figure}

For a disc with mixed magnetic fields, the expression for the
effective potential $V(r)$ is more complicated than the pure
toroidal field case. In Fig.~\ref{fig:potenm}, we show a sketch of $V(r)$
when both $B_z$ and $B_\phi$ are non-zero. We see that in addition to the
two MRs (IMR and OMR) (already present in the pure toroidal field case),
two \Alfven resonances (ARs) appear when $\tomega=\pm m\omega_{A\phi}$.
Since $m\omega_{A\phi}>\omega_s$, these ARs lie farther away from the
corotation radius than the MRs, at the distance
\be
r_{\rm ar}-r_{\rm c}=\pm\left({2\omega_{A\phi}\over 3\Omega}\right)r_{\rm c}.
\ee
Unlike the MRs, the ARs
are first-order singularities of the wave equation, i.e., $V(r)\propto
1/(\tomega-m\omega_{A\phi})\propto 1/(r-r_{\rm ar})$ near the ARs.

\section{Super-Reflection Of the Corotation Barrier}

As shown in Sec.~3, away from the CR/MRs/ARs region, the WKB dispersion
relation is simply
\be
\tomega^2=\kappa^2+k_r^2(c_s^2+v_{A}^2).
\ee
Thus waves can propagate either inside the ILR (where $\tomega=-\kappa$)
or outside the OLR (where $\tomega=\kappa$), and the region
between the ILR and OLR represents the ``corotation barrier'',
where waves are evanescent except for various possible singularities
associated with the CR/MRs/ARs.
We will be interested in the wave modes trapped between the disc inner
edge and ILR. As in the hydrodynamic ($B=0$) case, the growth rates of such p-modes depend directly
on the wave reflectivity of the corotation barrier (TL08, LT09).

Consider launching a density wave
$\delta h \propto \mbox{exp}(-{\rm i}\int^r k_r dr)$ from a radius inside
the ILR ($r<r_{\rm ILR}$) (assuming $k_r >0$; the minus sign arises from the fact that group velocity and phase velocity have different signs where $r<r_{\rm ILR}$, see TL08).
The wave carries negative energy since its pattern speed
$\omega/m$ is smaller than background rotation rate $\Omega$.
When the wave impinges on the corotation barrier, it is partly
bounced back as a reflection wave
$\delta h_{\rm R} \propto {\cal R}\,\mbox{exp}({\rm i}\int^r k_r dr)$,
which carries negative energy,
and partly transmitted through the evanescent zone and travels
beyond the OLR as a transmission wave
$\delta h_{\rm T} \propto {\cal T}  \mbox{exp}({\rm i}\int^r k_r dr)$,
which carries positive energy. ${\cal R}$ and ${\cal T}$ are reflection and transmission coefficients, respectively (see TL08).
Let the net absorption of wave energy at the CR, MRs and ARs be
${\cal D}_{\rm abs}$. Energy conservation then implies
\be
|{\cal R}|^2=1+|{\cal T}|^2+{\cal D_{\rm abs}}.
\label{eq:superf}
\ee
Obviously, wave transmission through the corotation barrier (the $|{\cal T}|^2$ term)
always tends to increase $|{\cal R}|^2$ beyond unity. However,
the resonant wave absorption term ${\cal D}_{\rm abs}$ can be either positive
or negative.

\subsection{Non-magnetic Discs}

TL08 showed that, for a $B=0$ barotropic disc, the wave super-reflection
$(|{\cal R}|^2-1)$ is dominated by the corotational wave absorption,
which depends on the dimensionless parameter
\be
\nu=\left({c_s\over p\kappa}{d\over dr}\ln\zeta\right)_{\rm c}
={2\over 3}\beta\left(\sigma-{3\over 2}\right),
\label{eq:nu}
\ee
where $\zeta$ is the vortensity [see Eq.~(\ref{eq:zeta})],
$\beta$, $p$ and $\sigma$ are defined by $\beta=c_s/(r\Omega)$,
$\Omega\propto r^{-p}$ and $\rho\propto r^{-\sigma}$ (around the CR),
respectively, and the subscript ``c'' implies that the quantity is evaluated at $r=r_{\rm c}$. The second equality in Eq.~(\ref{eq:nu}) assumes (Newtonian) Keplerian discs ($p=3/2$).
Importantly, the corotational wave absorption is proportional to
$\nu$ or $d\zeta/dr$. This can be understood from the fact that for
$\nu>0$ ($\nu<0$), the Rossby zone lies outside (inside) $r_{\rm c}$ (see Figs.~\ref{fig:poten1}-\ref{fig:poten3}),
implying a positive (negative) wave energy absorption.
Thus, if neglecting $|{\cal T}|^2$, we have
\be
{|\cal R|}-1\propto \nu\propto \left({d\zeta\over dr}\right)_{\rm c}.
\ee
Tsang \& Lai (2009b) further generalizes the analysis to non-barotropic
flows with radial stratification, in which case an effective
vortensity (which depends on the radial entropy profile of the disc)
plays a similar role as $\zeta$.


\subsection{Numerical calculations of reflectivity}
\label{subsec:refl}

For discs with magnetic fields, we are not able to
derive analytical expressions for the reflectivity. Thus we calculate
${\cal R}$ numerically.

For definiteness, we consider a (Newtonian) Keplerian disc with
\be
\Omega=\kappa\propto r^{-3/2},\quad \rho\propto r^{-\sigma},\quad
B_\phi \propto B_z \propto r^q,\quad {c_s\over r\Omega}=\beta,
\ee
where $\sigma$, $q$ and $\beta$ are constants. The strength of the
field is characterized by the ratio $b_\phi=v_{A\phi}/(r\Omega)$ and
$b_z=v_{Az}/(r\Omega)$, evaluated at $r_{\rm c}$.
We choose the wave frequency $\omega=\omega_r+{\rm i}\omega_i$, with
$0<\omega_i\ll \omega_r$.
The transmitted wave outside the OLR takes the form
\be
\delta h \propto A~\mbox{exp}\left({\rm i}\int^rk_r\,dr\right),
\ee
where $k_r>0$ is given by Eq.~(\ref{eq:kr}) or approximately
by Eq.~(\ref{eq:disp}), and $A(r)$ can be obtained from
Eq.~(\ref{eq:eta}) with $\eta\propto k_r^{-1/2}$.
Thus the boundary condition at $r_{\rm out} > r_{\rm OLR}$ is
\be
\frac{d\delta h}{dr}\big|_{r_{\rm out}}=\delta h
\left({\rm i}k_r +A'/A\right)\big|_{r_{\rm out}},
\label{eq:obc0}\ee
where $A'=dA/dr$.
Starting from $r=r_{\rm out}$, we integrate
Eqs.~(\ref{eq:ode1})-(\ref{eq:ode2}) inwards to a radius
inside the ILR. At $r=r_{\rm in}<r_{\rm ILR}$, the wavefunction
has the form
\be
\delta h \propto A~\left[\mbox{exp}\left(-{\rm i}\int^r k_r\,dr\right)
+{\cal R}\exp\left({\rm i}\int^r k_r\,dr\right)\right].
\ee
The reflectivity ${\cal R}$ can then be extracted from the equation
\be
|{\cal{R}}|=\left|\frac{(-{\rm i}k+A'/A)\delta h-\delta h'}{({\rm i}k+A'/A)\delta h
-\delta h'}\right|_{r_{\rm in}}.
\ee
In practice, at $r_{\rm in}$ ($r_{\rm out}$) sufficiently inside
$r_{\rm ILR}$ (outside $r_{\rm OLR}$), the $A'/A$ term can be dropped
since $k_r\gg |A'/A|$.

\begin{figure}
\begin{center}
\includegraphics[width=0.6\textwidth]{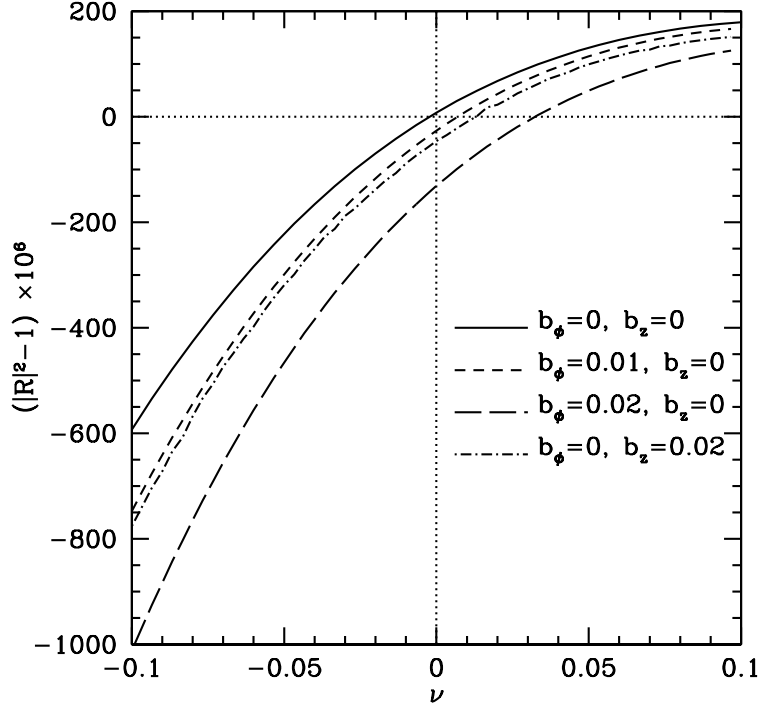}
\caption{Reflection coefficient as a function of $\nu$
for a Newtonian Keplerian disk. Different lines
represent different magnetic field strengths,
measured by the dimensionless parameters $b_\phi=(v_{A\phi}/r\Omega)_{\rm c}$
and $b_z=(v_{Az}/r\Omega)_{\rm c}$.
The other parameters are $m=2$, $B_\phi\propto B_z \propto r^q$ with $q=0$
and $\beta=c_s/(r\Omega)=0.1$. The parameter $\nu$ is defined
according to Eq.~(\ref{eq:nu}).}
\label{fig:reco}
\end{center}
\end{figure}

Figure \ref{fig:reco} shows the numerical results of the reflection coefficient
as a function of $\nu$, defined by Eq.~(\ref{eq:nu}).
For $B=0$, our result recovers that obtained by TL08.
We see even a relatively weak (sub-thermal, with $v_{A\phi}\lo c_s$) toroidal field
can significantly affect the reflectivity.
As the magnetic field increases, the reflectivity decreases
regardless of the sign of $\nu$. A finite vertical field has a similar effect, although it is sub-dominant compared to the toroidal field effect. We will show in the following section that this result
is consistent with the stability property of global disc p-modes.

\section{Global P-Modes of Black-Hole Accretion Discs}

In this section, we present our calculations of the global
non-axisymmetric p-modes in BH accretion discs. Recall that for a
nonmagnetic disc, these are the inertial-acoustic modes partially
trapped between the inner disc edge (at $r_{\rm in}=r_{\rm ISCO}$) and
the ILR. For a disc with finite magnetic fields, these become
inertial-fast magnetosonic waves.

We employ the standard shooting method (Press et al. 1992) to solve
Eqs.~(\ref{eq:ode1})-(\ref{eq:ode2}) for the complex eigenfrequency
$\omega=\omega_r+{\rm i}\omega_i$. The boundary condition at
$r_{\rm out}>r_{\rm OLR}$ is the outgoing-radiative condition
discussed in Sec.~\ref{subsec:refl} [see Eq.~(\ref{eq:obc0})].
At the inner boundary $r_{\rm in}=r_{\rm ISCO}$, we apply a free
surface boundary condition,
\be
\Delta h=\Delta \Pi/\rho=0,
\label{eq:inbc}\ee
i.e., the Lagrangian perturbation of the total pressure vanishes.
This boundary condition corresponds to a idealized situation in which
no wave energy is lost at the inner disc radius. As discussed in LT09 (and references therein),
the inner edge of a real BH disc is more complicated and may involve
wave energy loss due to radial infall of the accreting gas. Since our focus
here is to understand how disc magnetic fields affect the corotational
instability, we will adopt the simple boundary condition (\ref{eq:inbc})
for all the numerical calculations presented in this section.

As mentioned in Sec.~2, we use the Paczynski-Wiita pseudo-Newtonian
potential [Eq.~(\ref{eq:pwp})] to mimic the GR effect.
The disk density profile and magnetic field profile take the power-law forms
\be
\rho \propto r^{-\sigma}, \quad
B_{\phi}\propto B_z \propto r^{q}.
\ee
The thermal effect and the magnetic field effect are characterized
by the dimensionless parameters
\be
\beta=\frac{c_s}{r\Omega}, \quad
b_\phi=\frac{v_{A\phi}}{r\Omega}\big|_{r_{\rm in}},\quad
b_z=\frac{v_{Az}}{r\Omega}\big|_{r_{\rm in}},
\ee
with $\beta$ being constant.

The angular momentum flux carried by the wave across the magnetized
disc is given by (e.g. Pessah et al. 2006)
\be
F(r)=\pi r^2 \rho \mbox{Re}(\delta v_r\delta v_{\phi}^{\ast})
-\frac{1}{4}r^2\mbox{Re}(\delta B_r\delta B_{\phi}^{\ast}),
\label{eq:flux}\ee
where $\ast$ denotes complex conjugate. The first and second terms
in (\ref{eq:flux}) are related to the Reynolds stress and Maxwell stress,
respectively. The solution of Eqs.~(\ref{eq:ode1})-(\ref{eq:ode2})
gives $\xi_r(r)$ and $\delta h(r)$. The expressions for
$\delta v_r$, $\delta v_{\phi}$, $\delta B_r$
and $\delta B_{\phi}$ can be derived from the original linearized
perturbation equations (\ref{eq:mhd1})-(\ref{eq:mhd3}):
\be
\delta v_r=-i\tomega \xi_r,
\ee
\be
\delta v_{\phi}=\left(-\frac{\kappa^2}{2\Omega}
-\frac{m\omega_{A\phi}^2}{\tomega}
+\frac{mv_{A\phi}^2}{r\tomega}A_{11}\right)\xi_r
+\left(\frac{m}{r\tomega}+\frac{mv_{A\phi}^2}{r\tomega}A_{12}\right)
\delta h,
\ee
\be
\delta B_r=\frac{imB_{\phi}}{r}\xi_r,
\ee
\be
\delta B_{\phi}=\left(-q\frac{B_{\phi}}{r}-A_{11}B_{\phi}\right)
\xi_r -A_{12}B_{\phi}\delta h.
\ee

\subsection{Results for discs with pure toroidal magnetic fields}

\begin{figure}
\begin{center}$
\begin{array}{cc}
\includegraphics[width=0.45\textwidth]{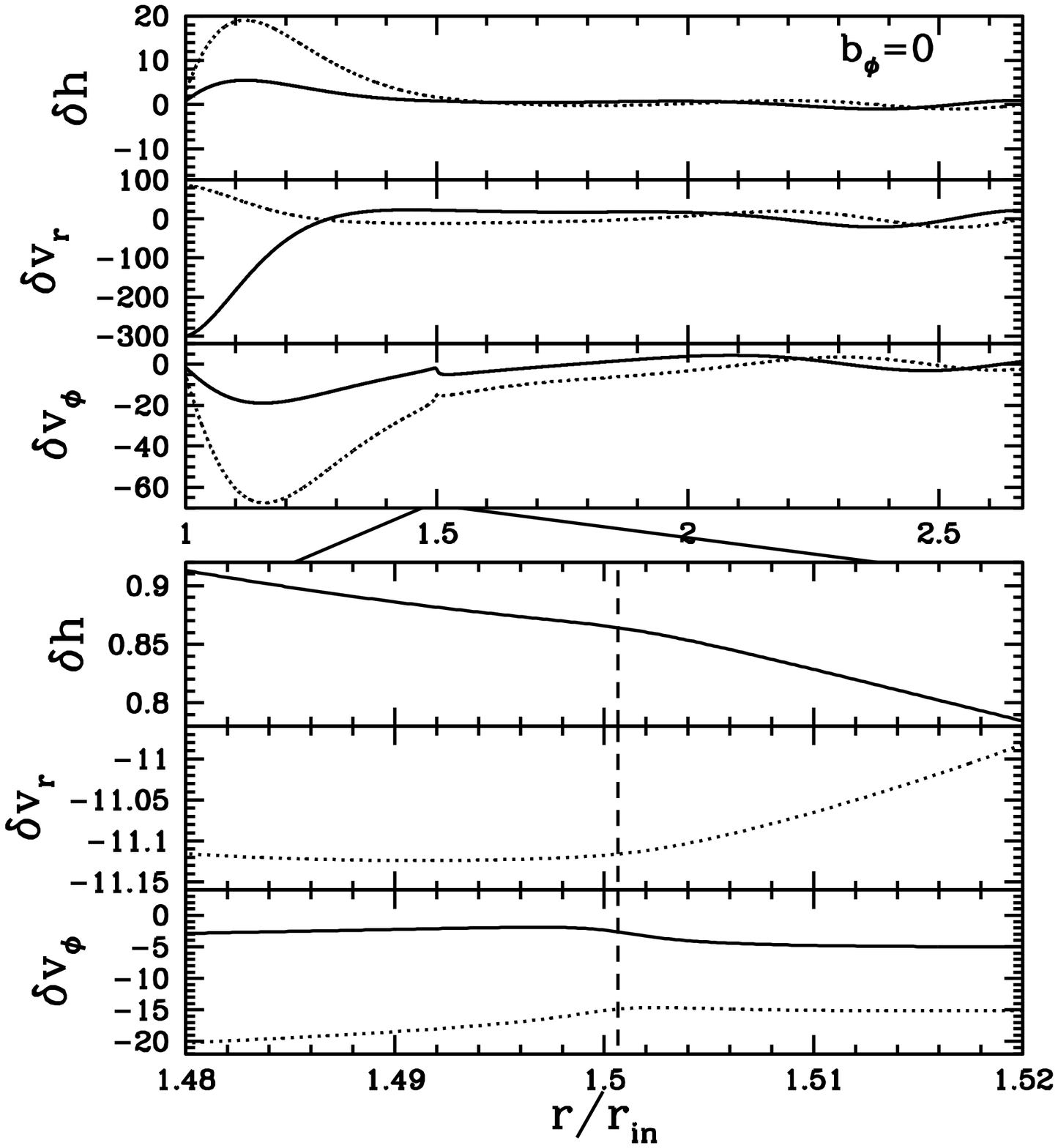}&
\includegraphics[width=0.45\textwidth]{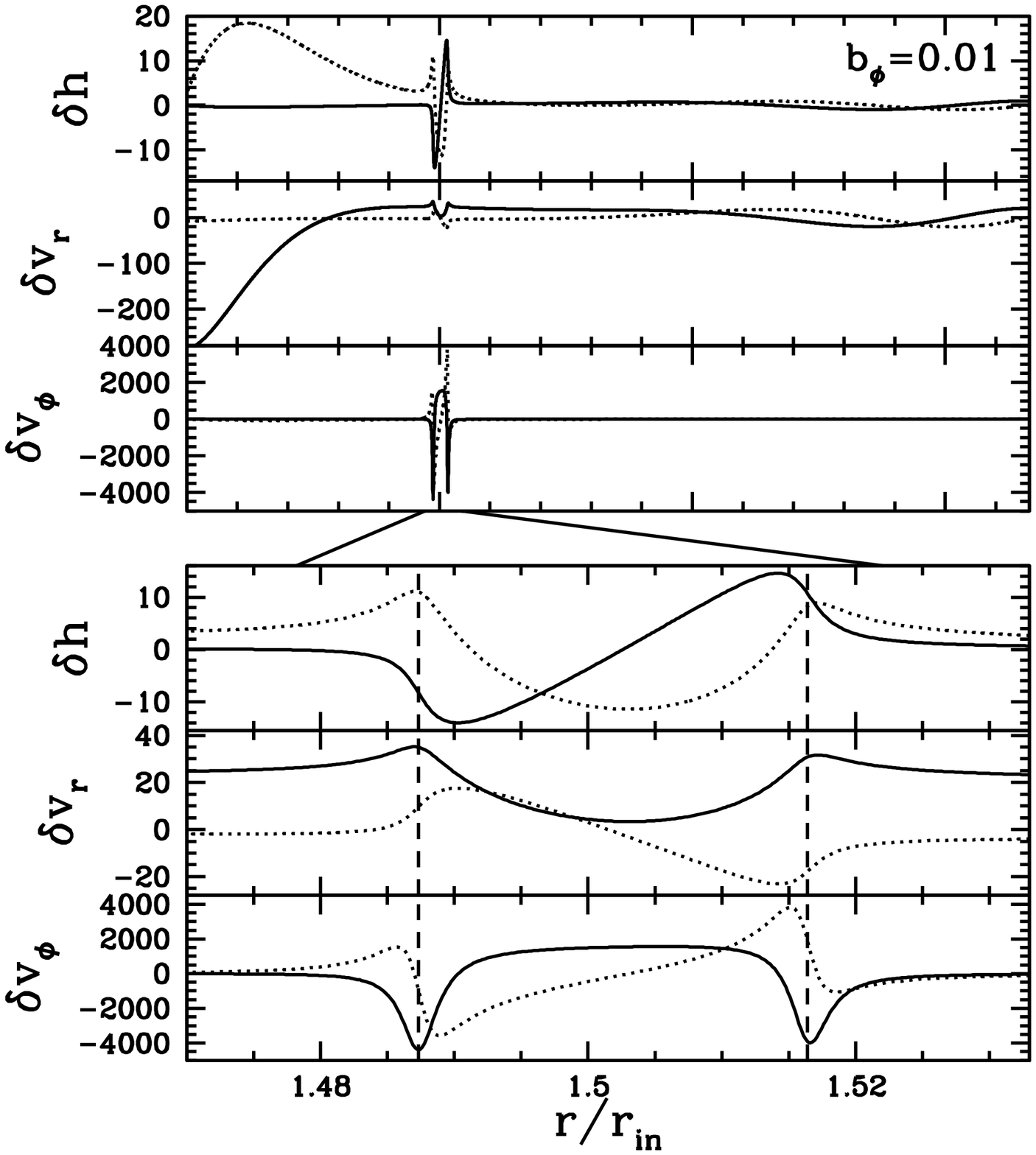}\\
&\includegraphics[width=0.45\textwidth]{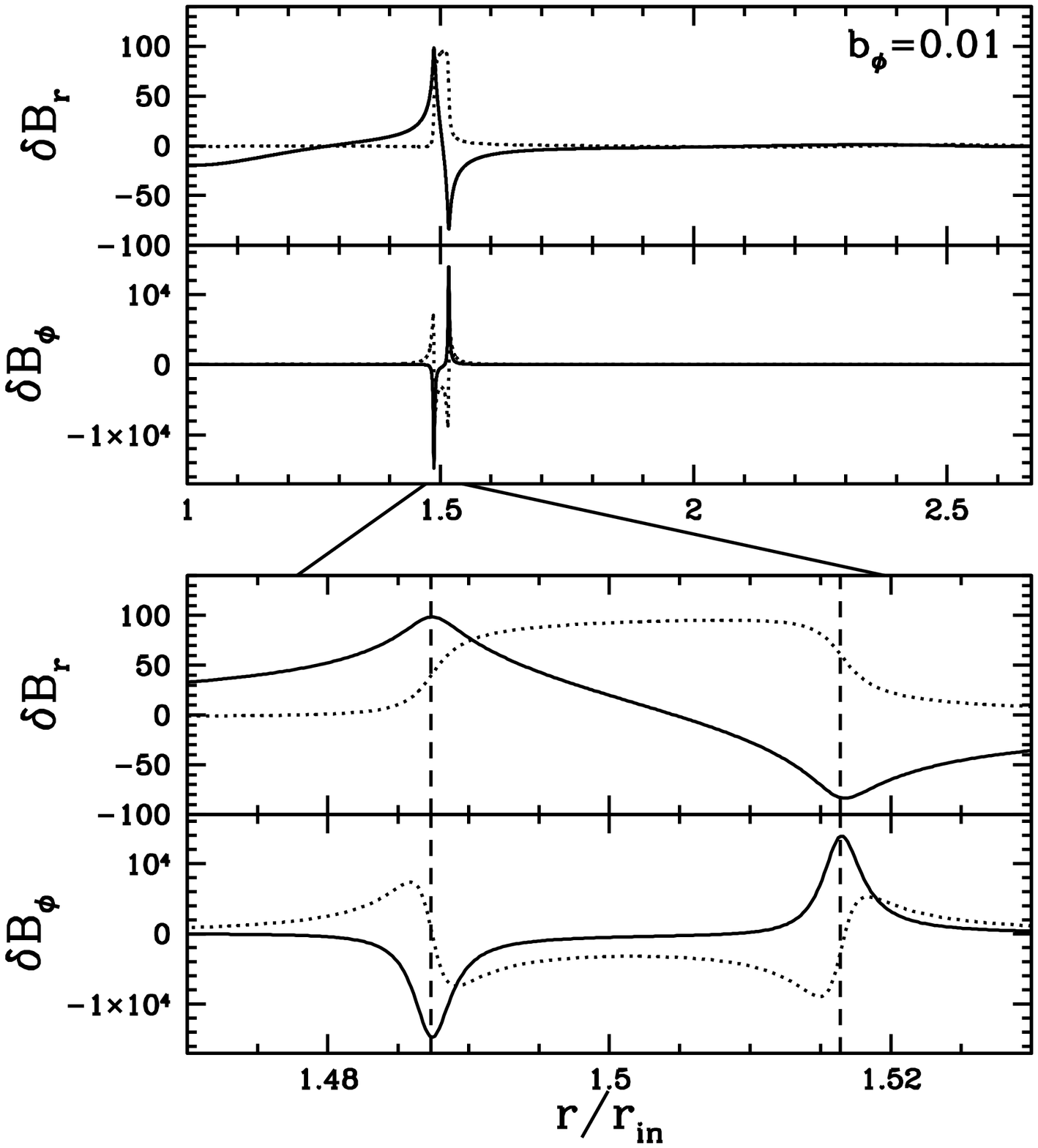}
\label{fig:wave}
\end{array}$
\end{center}
\caption{Example wavefunctions of $m=2$ p-modes in BH accretion discs.
  The disc density profile is $\rho\propto r^{-1}$ (i.e., $\sigma=1$)
  and the sound speed is $c_s=0.1r\Omega$ (i.e., $\beta=0.1$).
The left columns are for a $B=0$ disc, with the eigenvalue
$\omega=(0.9324+{\rm i}0.0027)\Omega_{\rm in}$ (where $\Omega_{\rm in}$ is
the disc rotation rate at $r=r_{\rm in}=r_{\rm ISCO}$); the right
panels are for a disc with $b_\phi=(v_{A\phi}/r\Omega)_{\rm in}=0.01$
[corresponding to $(v_{A\phi}/r\Omega)_{\rm c}=0.0175$]
(with $B_\phi$ independent of $r$, i.e., $q=0$), with $\omega
=(0.9312+i0.0018)\Omega_{\rm in}$.  In each subfigure, the upper panels show different variables as a function of $r$ in the full fluid zone. The
solid and dotted lines depict the real and imaginary parts,
respectively. The bottom panels zoom in the region near the corotation
resonance, with the vertical lines indicating the locations of the CR
(left) and magnetic slow resonances (right).  The vertical scales of
the wavefunctions are arbitrary, with $\delta h(r_{\rm out})=1$.}
\end{figure}

\begin{figure}
\begin{center}
\includegraphics[width=0.45\textwidth]{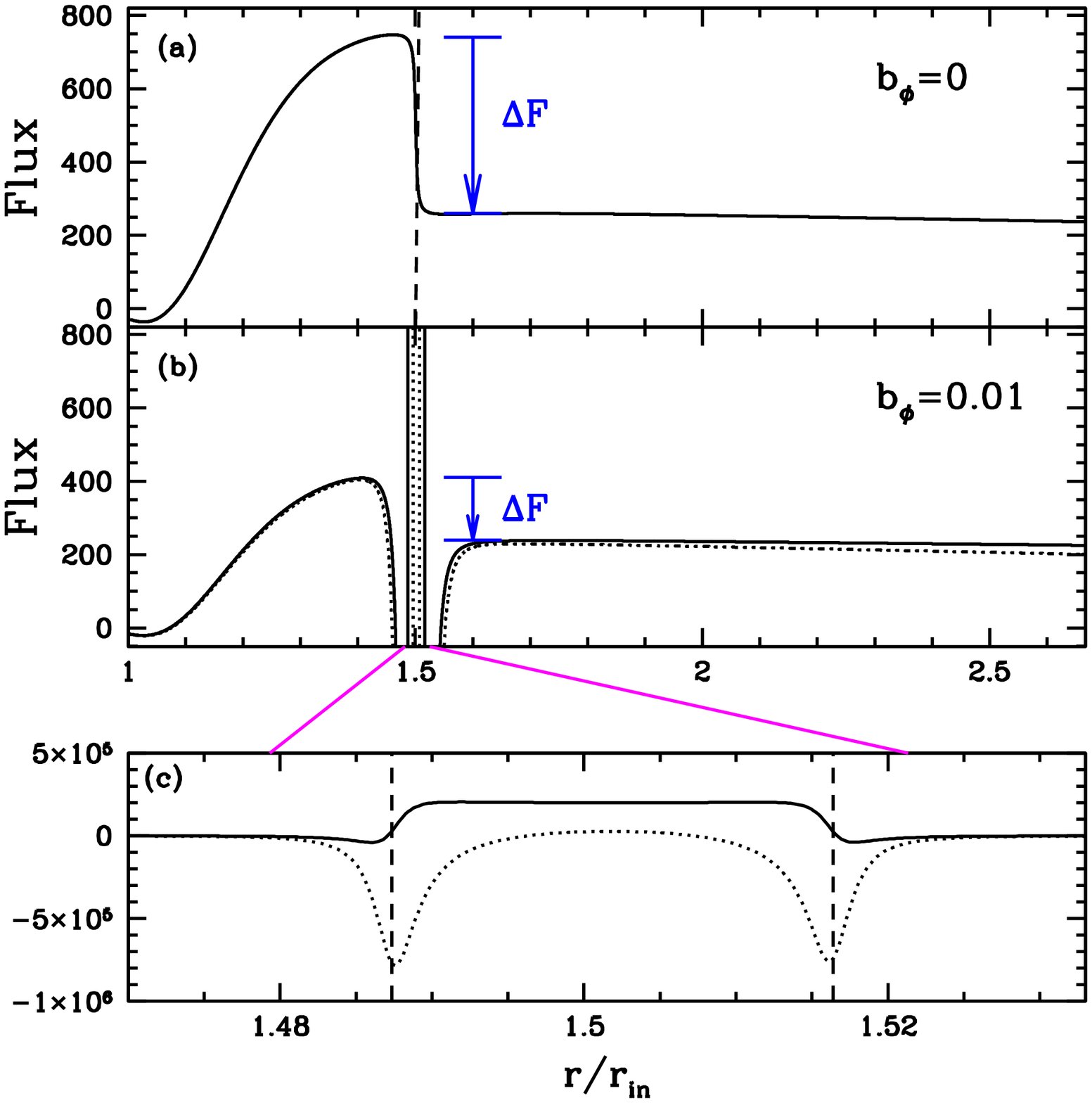}
\includegraphics[width=0.45\textwidth]{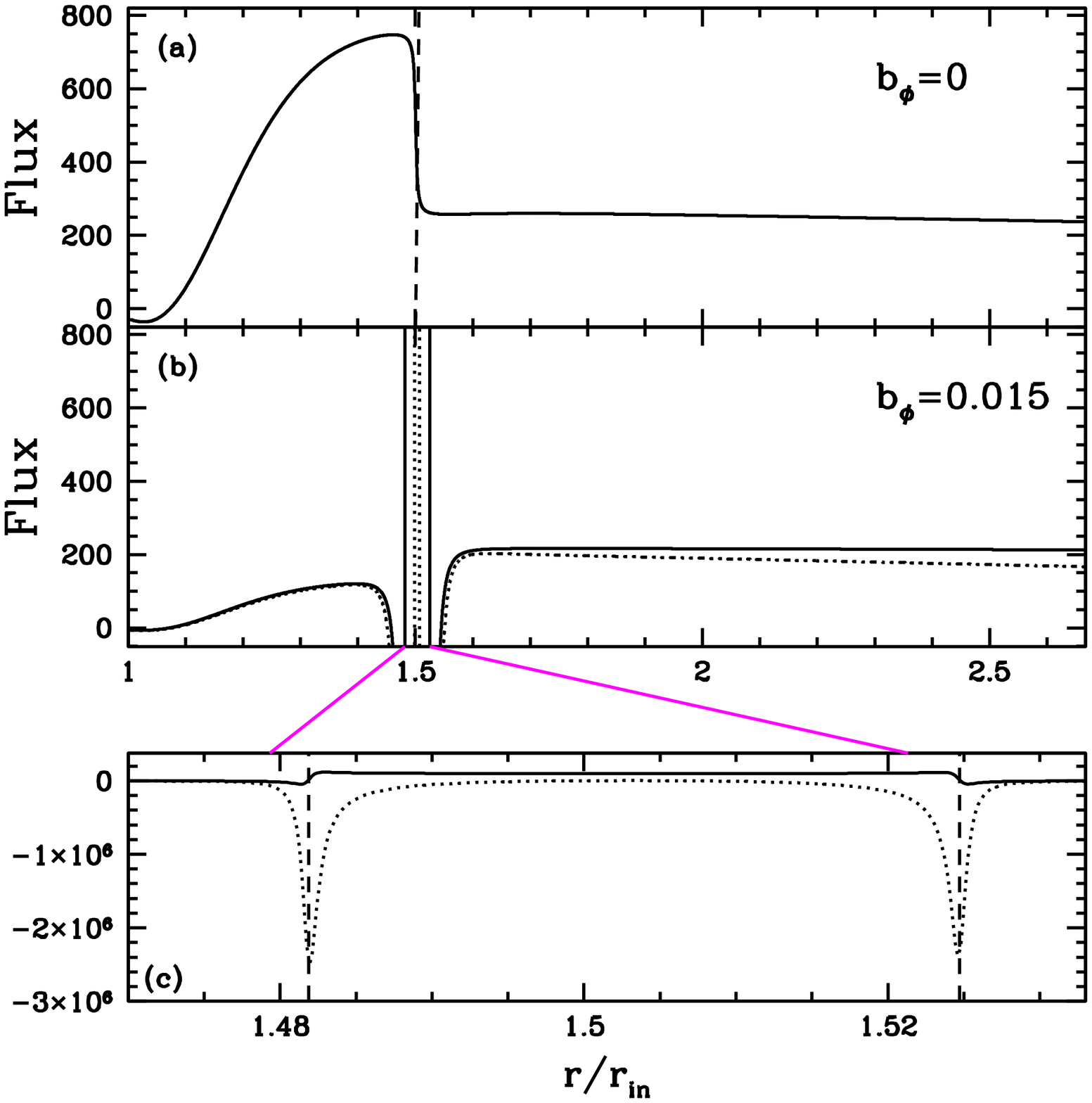}
\caption{Angular momentum flux carried by the wave as a function of
  $r$. The disc parameters are $\sigma=1$, $q=0$ and $\beta=0.1$, and
  the wave modes have $m=2$. Panel (a) shows the result of a $B=0$
  disc [with the mode frequency $\omega=(0.9324+{\rm i}0.0027)\Omega_{\rm
      in}$] with the vertical line indicating the corotation
  resonance. Panel (b) is for a magnetized disc with $b_\phi=0.01$
[corresponding to $(v_{A\phi}/r\Omega)_{\rm c}=0.0175$] [left column;
    with $\omega=(0.9312+{\rm i}0.0018)\Omega_{\rm in}$] or $b_\phi=0.015$
[corresponding to $(v_{A\phi}/r\Omega)_{\rm c}=0.026)$] [right column;
    with $\omega=(0.93+{\rm i}0.00065)\Omega_{\rm in}$]. The solid and
  dotted curves show the total angular momentum flux and the flux
  carried by the fluid motion only (the Reynolds stress),
  respectively. The bottom panels are the blow-up of panel (b) near
  the magnetic resonances whose locations are indicated by the two
  vertical lines.}
\label{fig:flux}
\end{center}
\end{figure}

\begin{figure}
\begin{center}
\includegraphics[width=0.6\textwidth]{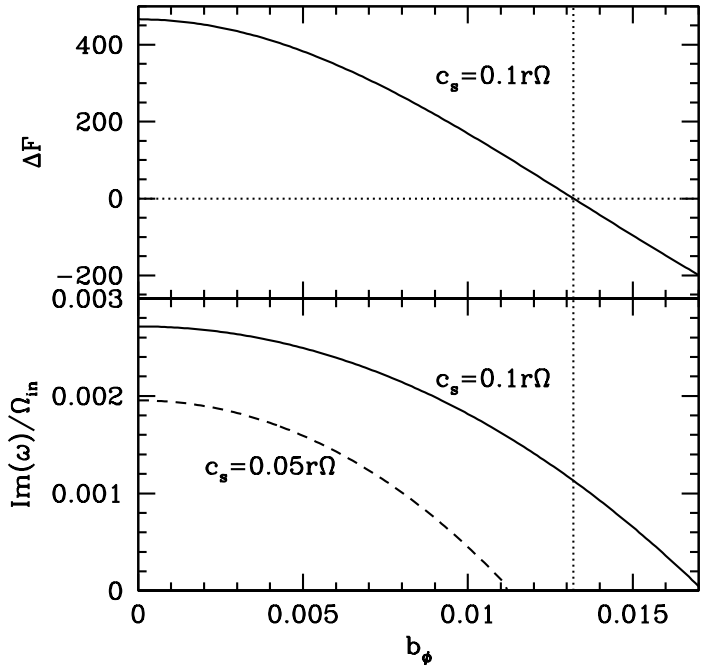}
\caption{The net angular momentum flux jump across the two
magnetic resonances (upper panel) and the mode growth rate (lower panel)
as a function of the dimensionless toroidal magnetic field strength
$b_\phi=(v_{A\phi}/r\Omega)_{\rm in}$. The solid lines are for disc sound speed $c_s=0.1r\Omega$, and the dashed line in the lower panel is for $c_s=0.05r\Omega$.
The other parameters are the same as in Figs.~\ref{fig:wave} and \ref{fig:flux}:
$m=2$, $\sigma=1$, $q=0$.}
\label{fig:fluxd}
\end{center}
\end{figure}

\begin{figure}
\begin{center}
\includegraphics[width=0.6\textwidth]{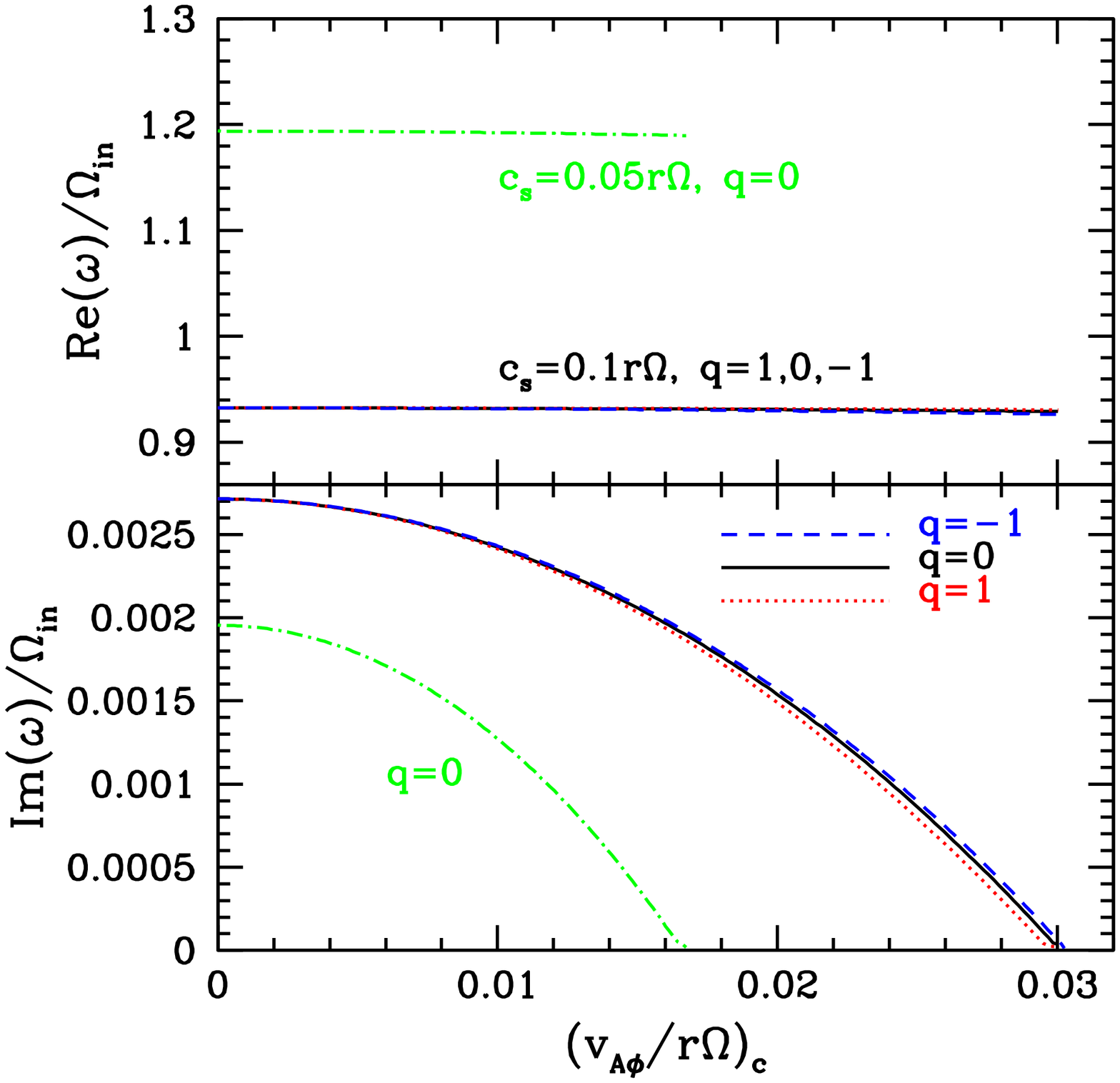}
\caption{Frequencies of the $m=2$ p-modes as a function of the
  dimensionless toroidal magnetic field strength,
  $(v_{A\phi}/r\Omega)_{\rm c}$, with the upper and lower panels
  showing the real and imaginary parts, respectively. Different curves
  correspond to different values of $q=d\ln B_{\phi}/d\ln r$ (i.e.,
  the slope of $B_{\phi}$ profile) and sound speed $c_s$.  The disc
  density profile is $\rho \propto r^{-1}$, i.e. $\sigma=1$.}
\label{fig:grp}
\end{center}
\end{figure}

Figure \ref{fig:wave} gives an example of the eigenfunctions for
overstable p-modes in magnetized BH accretion discs.
For comparison, the corresponding results for a $B=0$ disc are also
shown.  Clearly, with a small B field ($b_\phi\ll 1$), the
wavefunctions away from the corotation/magnetic resonances (CR/MRs)
are only slightly modified. The wave action
is still trapped in the inner disc region.
The most noticeable effects occur near the MRs, where both $\delta h$
and $\delta v_r$ experience significant variations.  In this example,
the $B=0$ disc has $d\zeta/dr>0$ at the CR, so that corotational wave
absorption leads to a mode growth rate $\omega_i\simeq
0.0027\Omega_{\rm in}$.  The inclusion of a small toroidal B
field, $b_\phi=0.01$
[corresponding to $(v_{A\phi}/r\Omega)_{\rm c}=0.0175$],
however, reduces the mode growth rate to
$\omega_i\simeq 0.0018\Omega_{\rm in}$. Note that the disc sound speed in the example is $c_s=0.1r\Omega$, so the $b_\phi=0.01$ corresponds to $(v_{A\phi}/c_s)_{\rm c}=0.175$.

The origin of the diminished mode growth rate due to magnetic fields
can be understood by examining the angular momentum flux carried by the
wave modes. Figure~\ref{fig:flux} gives some examples. We see that in
the $B=0$ case, there is a jump of the angular momentum flux
across the CR,
\be
\Delta F=F(r_{\rm c-})-F(r_{\rm c+}),\qquad (B=0).
\ee
The fact that $\Delta F>0$ implies a positive wave energy (angular momentum)
absorbed at the CR [i.e., ${\cal D}_{\rm abs}>0$ in Eq.~(\ref{eq:superf})];
this leads to super-reflection and
is the main driver for the overstability of hydrodynamic
p-modes. For discs with a finite $B_\phi$, the CR is split into two MRs
(the IMR and OMR), and we see from Fig.~\ref{fig:flux} that significant flux
jumps occur at both the IMR and OMR. The flux jump at
the IMR, $\Delta F_{\rm IMR}=F(r_{\rm IMR-})-F(r_{\rm IMR+})$, is negative,
implying a negative angular momentum absorption. The flux jump at
the OMR, $\Delta F_{\rm OMR}=F(r_{\rm OMR-})-F(r_{\rm OMR+})$, however,
is positive, implying a positive angular momentum absorption.
The net angular momentum jump across the MRs region is then
\be
\Delta F=F(r_{\rm IMR-})-F(r_{\rm OMR+})=\Delta F_{\rm IMR}+\Delta F_{\rm OMR}.
\ee
Note that although both $|\Delta F_{\rm IMR}|$ and $|\Delta F_{\rm OMR}|$ are
much larger (by orders of magnitude) than $\Delta F$ in the $B=0$ case,
the net jump $\Delta F$ is smaller for discs with finite $B_\phi$.
As $B_\phi$ increases, $|\Delta F_{\rm IMR}|$ and $|\Delta F_{\rm OMR}|$ both
decrease, and $\Delta F$ becomes smaller and may even change sign (See
Fig.~\ref{fig:fluxd}). This decrease of wave absorption at the MRs
due to finite magnetic fields results in a reduced or even diminished
super-reflection (see Sec.~4), leading to a reduction of the growth rate and
even stabilization of disc p-modes.

Figure \ref{fig:fluxd} shows the p-mode growth rate as a function of
$b_\phi$ and the corresponding net angular momentum jump across the
CR/MRs region. Obviously, the decrease of $\Delta F$ with increasing
$b_\phi$ directly correlates with the decrease of the mode growth
rate. Note that when $\Delta F$ is zero or slightly negative, the
p-mode still has a finite growth rate. This is because wave
transmission across the corotation barrier always tends to increase
the reflectivity [see Eq.~(\ref{eq:superf})], thereby promoting the
the mode growth. With a further increase of $b_\phi$, the mode growth
becomes completely suppressed. For example, for discs with
$c_s=0.1r\Omega$, the angular momentum absorption at the CR/MRs change
sign at $b_\phi=0.0132$, while the mode growth rate becomes zero at
$b_\phi \simeq 0.017$ , or
$(v_{A\phi}/c_s)_{\rm in} \simeq 0.17$
[corresponding to $(v_{A\phi}/c_s)_{\rm c}=0.3$].
For discs with $c_s=0.05r\Omega$, the mode growth rate vanishes at $b_\phi
\simeq0.01$, or
$(v_{A\phi}/c_s)_{\rm in} \simeq 0.2$
[corresponding to $(v_{A\phi}/c_s)_{\rm c}=0.34$].

Figure \ref{fig:grp} shows the real and imaginary parts of the $m=2$ mode
frequency as a function of the dimensionless field strength
$v_{A\phi}/(r\Omega)$ (evaluated at $r_c$) for different disc sound speeds
and different values of $q$ (measuring the slope of $B_\phi$).
Clearly, for such a small field strength, the real mode frequency
is approximately independent of $B_\phi$ and $q$, since the propagation zone
of the wave mode is hardly modified by the magnetic field.
The mode growth rate, however, is significantly affected because
of the modification to the CR. Not surprisingly,
${\rm Im}(\omega)$ depends on $B_\phi$ mainly through the ratio
$(v_{A\phi}/r\Omega)_{\rm c}$.

\begin{figure}
\begin{center}
\includegraphics[width=0.5\textwidth]{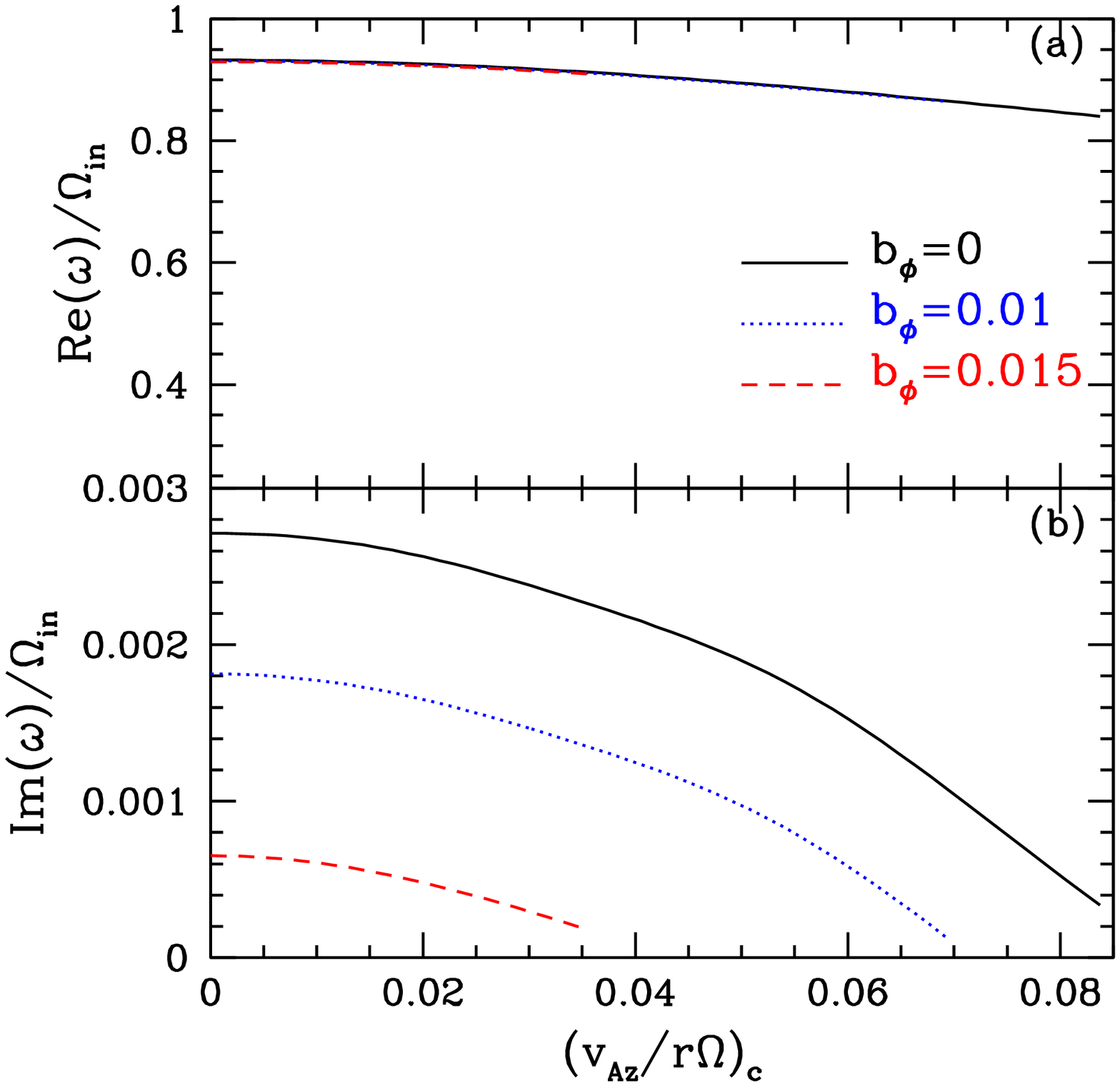}
\caption{Frequencies of the $m=2$ p-modes as a function of the
  dimensionless vertical magnetic field strength,
  $(v_{Az}/r\Omega)_{\rm c}$, with the upper and lower panels
  showing the real and imaginary parts, respectively. Different curves
  correspond to different values of the disc toroidal fields. Note that the real mode
frequency depends weakly on the magnetic field (for the range of
field strength considered), thus $b_\phi=0.01$ corresponds to $(v_{A\phi}/r\Omega)_{\rm c}\simeq 0.0175$ and $b_\phi=0.015$ corresponds to $(v_{A\phi}/r\Omega)_{\rm c}\simeq 0.026$. The other disc parameters are: $\sigma=1$, $q=0$ and $c_s=0.1r\Omega$.}
\label{fig:grm}
\end{center}
\end{figure}

\begin{figure}
\begin{center}
\includegraphics[width=0.6\textwidth]{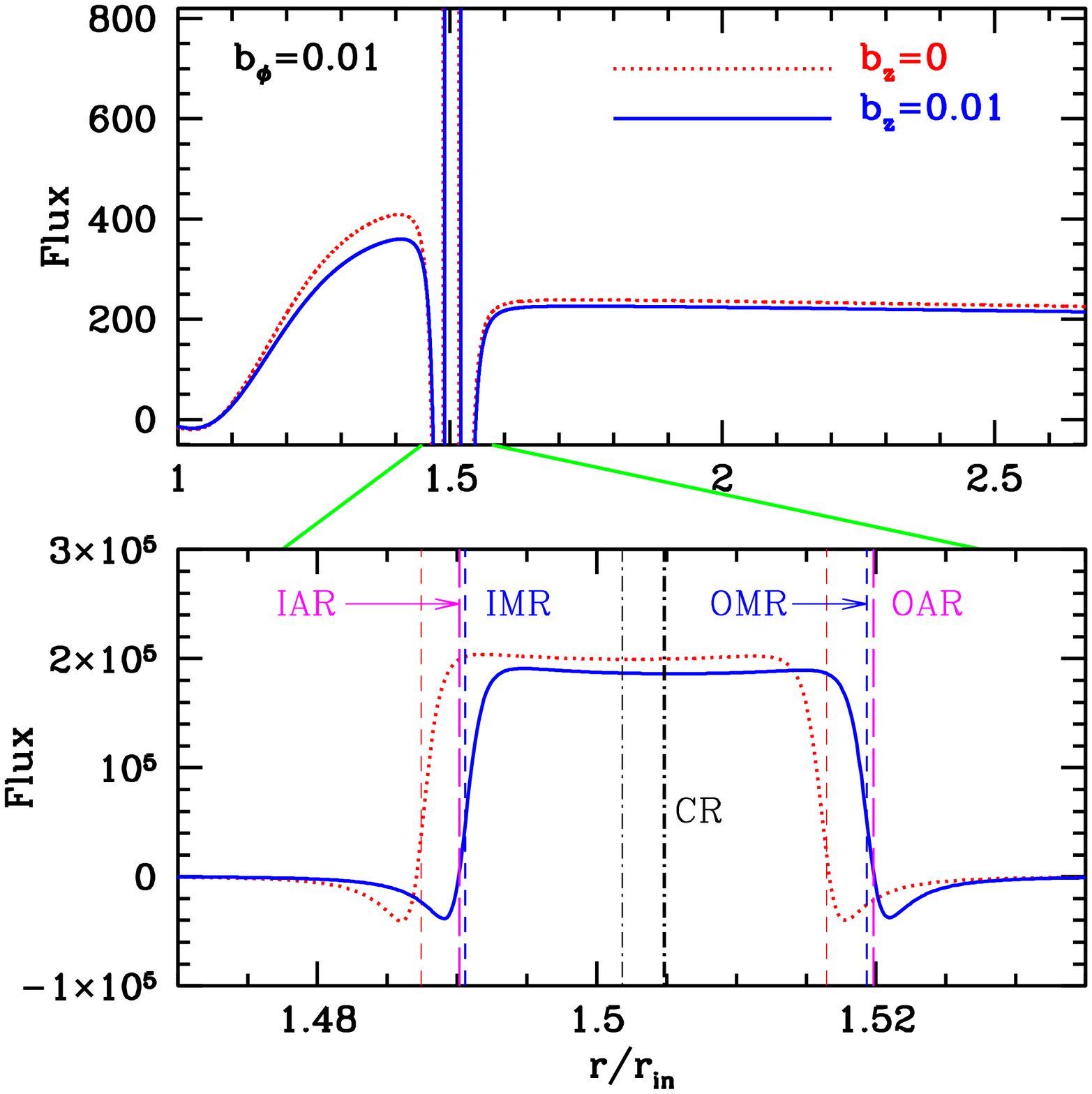}
\caption{The angular momentum flux carried by wave modes as a function of
radius in a BH accretion disc with mixed magnetic fields. The bottom panel is
the zoom-in of the upper panel near the resonances.
The dotted curves are for a disc with a pure toroidal field, while the
solid curves are for a disc with both toroidal and vertical fields.
The short-dashed vertical lines
indicate the inner and outer magnetic slow resonances (IMR and OMR).
The dot-dashed vertical lines represent
the corotation resonance (CR). The long-dashed vertical lines
(very close to the short-dash lines) indicate the inner/outer \Alfven resonances (IAR and OAR) which
only exist in the mixed field case. Note that the mode frequency in the
$b_z=0$ case is different from the $b_z=0.01$ case, thus there is a
slight shift in location of the MRs and CR between the two cases.
Note that $b_\phi=0.01$ and $b_z=0.01$ correspond to
$(v_{A\phi}/r\Omega)_{\rm c}=(v_{Az}/r\Omega)_{\rm c}=0.0176$.
The other disc parameters are: $\sigma=1$, $q=0$ and $c_s=0.1r\Omega$.}
\label{fig:fm}
\end{center}
\end{figure}

\subsection{Results for discs with mixed magnetic fields}

The effects of vertical magnetic fields and mixed fields on the disc
p-mode growth rate are illustrated in Fig.~\ref{fig:grm} and
Fig.~\ref{fig:fm}.  We see that the real mode frequency depends very
weakly on the field strength unless $v_A$ becomes comparable to or
larger than $c_s$. The solid line in Fig.~\ref{fig:grm}b shows that
for the pure vertical field case, the p-mode growth rate decreases
with increasing $B_z$.  This is consistent with our finding in
Sec.~4.2 that a finite $B_z$ tends to reduce the reflectivity (see
Fig.~\ref{fig:reco}). When the toroidal field is also included, we see
the similar trend (see the dotted and dashed lines). Thus both
vertical and toroidal magnetic fields tend to suppress the
corotational instability.  The effect of the toroidal magnetic field
is somewhat larger that the vertical field, given that the required
toroidal field strength $B_{\phi}$ to completely suppress the mode
growth is smaller than the required $B_z$. This is understandable
since the p-modes have no vertical structure (with $k_z=0$), and fluid
motion in the disc does not directly bend the vertical field lines.

In Fig.~\ref{fig:fm}, we compare the angular momentum flux profiles for
the pure toroidal field case and the mixed field case. From the
upper panel of this figure we see that for the mixed field case, the
net flux jump across the resonance zone is smaller than the pure
toroidal field case. This explains the somewhat smaller mode growth
rate ($\omega_i/\Omega_{\rm in}=0.0017$) for the mixed field case as
compared to the pure toroidal field case ($\omega_i/\Omega_{\rm in}=0.0018$). Note that the real part of the eigenfrequency is also
modified slightly by the vertical field.  This gives rise to the shift of
``plateau'' in the bottom panel of Fig.~\ref{fig:fm}.

\section{Conclusion}

In this paper we have investigated the effects of magnetic fields on
the corotational instability of the p-modes (also called
inertial-acoustic modes for the $B=0$ case) in BH accretion
discs. These modes have no vertical structure, are trapped between the
inner disc edge and the inner Lindblad resonance, and have the
character of inertial-fast magnetosonic waves in their propagation
zone. Previous works have shown that in a hydrodynamic disc (with
$B=0$) these modes can be overstable due to wave absorption at the
corotation resonance (CR) when the vortensity (or its generalization
for non-barotropic flows) of the disc has a positive gradient (TL08,
LT09, Tsang \& Lai 2009b). In this paper we found that the inclusion
of a finite toroidal field $B_{\phi}$ splits the CR into two magnetic
(slow) resonances (MRs), where the wave frequency in the rotating
frame of the fluid, $\omega-m\Omega$, matches the slow magnetosonic
wave frequency.  At the inner and outer MRs, the angular momentum flux
carried by the wave undergoes significant change, indicating angular
momentum (and energy) absorption at both MRs. One of the resonances
absorbs positive angular momentum while the other emits positive
angular momentum. Independent of the background flow vortensity
gradient, the net angular momentum absorption across the resonance
region is always reduced or becomes more negative in discs with
$B_\phi\neq 0$ compared to the $B=0$ case.  This leads to a reduced
growth rate of the p-modes.  Our calculations showed that the
hydrodynamically overstable inertial-acoustic modes can be completely
stabilized by the toroidal field at relatively small field
strengths (see Fig.~\ref{fig:grp})
For example, for discs with sound speed $c_s=0.1r\Omega$, the $m=2$
p-mode growth rate vanishes at
$v_{A\phi}/(r\Omega)\sim 0.03$ (evaluated at the corotation radius $r_c$;
where $v_{A\phi}=B_\phi/\sqrt{4\pi\rho}$ is the \Alfven speed
associated with $B_\phi$),
corresponding to $(v_{A\phi}/c_s)_{\rm c}\sim 0.3$.
We also carried out numerical calculations
of the reflectivity of a wave approaching the corotation from small
radii.  We showed that the reflectivity is reduced by the toroidal
magnetic field, in agreement with our global mode calculation results.

For discs with a pure vertical field, the corotation resonance persists for the p-modes,
with no additional magnetic resonances. This is understandable since
p-modes have no vertical structure ($k_z=0$).  With no bending of
the field lines, the vertical field simply modifies the background
pressure, or equivalently, changes the effective background sound
speed. However, when $B_z\neq 0$, the effective potential for wave
propagation contains a second-order singularity term, in addition to
the first-order singularity already present in a $B_z=0$ disc.

For discs containing mixed (toroidal and vertical) magnetic fields,
the corotation resonance is split into four resonances: in addition to the inner/outer
MRs (already present in discs with pure toroidal fields), two \Alfven
resonances appear, where $\omega-m\Omega$ matches the local \Alfven
wave frequency. We showed that the effect of these additional
resonances is to further reduce the super-reflection and the growth
rate of the disc p-modes
(see Fig.~\ref{fig:grm}).
Overall, the toroidal field has a larger effect
on the corotational wave absorption and the mode growth rate than the
vertical field.

The main finding of our paper is that overstable non-axisymmetric
p-modes tend to be stabilized by disc magnetic fields. This implies
that, in order for p-modes to explain HFQPOs observed in black-hole X-ray
binaries, the disc magnetic field must be sufficiently weak (with the
magnetic pressure appreciably less than the thermal pressure) so that
the corotational instability is not completely
suppressed. Alternatively, some other mode excitation mechanisms are
needed. One possibility could be mode excitations by turbulent
viscosity (see Kato 2001 for a general discussion of viscous driving
of disc oscillation modes) or by vorticity perturbations associated
with the turbulence (see Heinemann \& Papaloizou 2009).  Another
possibility could be the accretion-ejection instability studied by
Tagger et al.~(see Tagger \& Pellat 1999; Varniere \& Tagger 2002;
Tagger \& Varniere 2006), which involves large-scale equipartition
vertical magnetic fields threading a thin disc embedded in a vacuum or a
tenuous corona.  Whether these other possible excitation mechanisms
can compete with the stabilizing effects of the internal disc magnetic
fields remains to be studied. Nevertheless, our finding is limited by the fact that the disc model we considered is essentially an infinite cylinder without vertical structure, whereas in reality disc background (density, magnetic field, etc.) varies in vertical direction. More realistic disc models certainly need to be investigated in future studies.

Finally, we note that although in this paper we have studied the role
of magnetic fields in the context of black-hole diskoseismic oscillations and
QPOs, the dynamical effects of magnetic fields on the corotation
resonance examined here are quite general. Our finding in this paper
suggests that dynamical instabilities in other rotating astrophysical
flows where the corotation resonance plays an important role may be
significantly affected by magnetic fields. Examples include the
Papaloizou-Pringle instability in accretion tori and the global
rotational instability in differentially rotating stars. We will study
some of these issues in separate papers (Fu \& Lai 2010).

\section*{Acknowledgments}

We thank Thierry Foglizzo, Michel Tagger and especially David Tsang
for useful discussion during the course of this study. We are grateful to an anonymous referee whose comments helped improve this manuscript. This work has
been supported in part by NASA Grant NNX07AG81G and NSF grants AST
0707628. DL also acknowledges the hospitality (Jan -- Jun, 2010) of the Kavli
Institute for Theoretical Physics at UCSB, funded by the NSF through
grant PHY05-51164.


\label{lastpage}
\end{document}